\documentclass[camera,10pt]{jpaper} 
\DeclareMathAlphabet{\mathcal}{OMS}{cmsy}{m}{n}
\usepackage{setspace} 
\usepackage[italic]{mathastext}
\usepackage{array}
\usepackage{balance}
\newcommand{\ignore}[1]{}
\usepackage{fancyhdr}
\usepackage[normalem]{ulem}
\usepackage{datetime} 
\usepackage{siunitx}
\usepackage{booktabs}
\usepackage{dblfloatfix}
\usepackage{hyperref}
\usepackage{multirow}
\usepackage{multicol}
\usepackage{float}
\usepackage[shortlabels]{enumitem}
\usepackage{xurl}
\usepackage{cite}
\usepackage[resetlabels,labeled]{multibib}
\usepackage{xcolor}
\usepackage[font={small,bf}]{caption}
\usepackage{pifont}
\usepackage{amssymb}
\usepackage{amsmath}

\newcommand{\cmark}{\ding{51}}%

\definecolor{dollarbill}{rgb}{0.52, 0.73, 0.4}

\newcolumntype{L}[1]{>{\raggedright\let\newline\\\arraybackslash\hspace{0pt}}m{#1}}
\newcolumntype{C}[1]{>{\centering\let\newline\\\arraybackslash\hspace{0pt}}m{#1}}
\newcolumntype{R}[1]{>{\raggedleft\let\newline\\\arraybackslash\hspace{0pt}}m{#1}}

\setlist[itemize]{noitemsep,itemsep=0pt,parsep=0pt,topsep=0pt,partopsep=0pt,leftmargin=1em}
\setlist[enumerate]{noitemsep,itemsep=0pt,parsep=0pt,topsep=0pt,partopsep=0pt,leftmargin=1em}



\definecolor{grey}{rgb}{0.7, 0.6, 0.6}
\definecolor{amber}{rgb}{1.0, 0.49, 0.0}
\definecolor{darkamber}{rgb}{0.9, 0.49, 0.0}
\definecolor{extradarkamber}{rgb}{0.55, 0.295, 0.0}
\definecolor{darkgreen}{rgb}{0.0, 0.2, 0.13}
\definecolor{darkbyzantium}{rgb}{0.36, 0.22, 0.33}
\definecolor{darkseagreen}{rgb}{0.56, 0.74, 0.56}
\definecolor{darkspringgreen}{rgb}{0.09, 0.45, 0.27}
\definecolor{dollarbill}{rgb}{0.52, 0.73, 0.4}
\definecolor{darkcerulean}{rgb}{0.03, 0.27, 0.49}

\newif\ifcameraready
\camerareadytrue

\ifcameraready
  \newcommand{\xmp}[1]{#1}
  \newcommand{\xmt}[1]{#1}
  
  \newcommand{\xmg}[1]{#1}
\else
\newcommand{\xmp}[1]{#1}
\newcommand{\xmt}[1]{#1}

\newcommand{\xmg}[1]{#1}
\fi

\newcommand{\mpo}[1]{#1}
\newcommand{\mpt}[1]{#1}
\newcommand{\mph}[1]{#1}
\newcommand{\mpf}[1]{#1}
\newcommand{\mpg}[1]{#1}
\newcommand{\mpi}[1]{#1}
\newcommand{\mpj}[1]{#1}
\newcommand{\mpk}[1]{#1}
\newcommand{\mpl}[1]{#1}
\newcommand{\mpm}[1]{#1}
\newcommand{\mpn}[1]{#1}

\ifcameraready
  \usepackage[disable]{todonotes}
   
  \else
  \usepackage[textsize=scriptsize]{todonotes}
   
\fi

\makeatletter
\g@addto@macro{\normalsize}{%
  \setlength{\abovedisplayskip}{2pt plus 1pt minus 1pt}
  \setlength{\belowdisplayskip}{2pt plus 1pt minus 1pt}
  \setlength{\abovedisplayshortskip}{0pt}
  \setlength{\belowdisplayshortskip}{0pt}
  \setlength{\intextsep}{4pt plus 1pt minus 1pt}
  \setlength{\textfloatsep}{4pt plus 1pt minus 1pt}
  \setlength{\skip\footins}{5pt plus 1pt minus 1pt}}
\makeatother

\titlespacing\section{0pt}{2pt plus 1pt minus 1pt}{3pt plus 1pt minus 2pt}
\titlespacing\subsection{0pt}{2pt plus 1pt minus 1pt}{3pt plus 1pt minus 2pt}
\titlespacing\subsubsection{0pt}{2pt plus 1pt minus 1pt}{3pt plus 1pt minus 2pt}

\newcommand*\circled[1]{\tikz[baseline=(char.base)]{
    \node[shape=circle,fill,inner sep=1pt] (char) {\textcolor{white}{\textbf{#1}}};}}

\usepackage{tikz}
\usepackage{fancyhdr}
\usepackage{datetime} 

\settimeformat{ampmtime}

\newif\ifuseversions
\useversionsfalse
\newcommand{\versionnum}[0]{26}
\ifuseversions
  \def\parsepdfdatetime#1:#2#3#4#5#6#7#8#9{%
    \def\theyear{#2#3#4#5}%
    \def\themonth{#6#7}%
    \def\theday{#8#9}%
    \parsepdftime
  }

  \def\parsepdftime#1#2#3#4#5#6#7\endparsepdfdatetime{%
    \def\thehour{#1#2}%
    \def\theminute{#3#4}%
    \def\thesecond{#5#6}%
    \ifstrequal{#7}{Z}
    {%
      \def\thetimezonehour{+00}%
      \def\thetimezoneminute{00}%
    }%
    {%
      \parsepdftimezone#7%
    }%
  }

  \def\parsepdftimezone#1'#2'{%
    \def\thetimezonehour{#1}%
    \def\thetimezoneminute{#2}%
  }

  \newcommand*{\thetimezone}{\thetimezonehour:\thetimezoneminute}
  \expandafter\parsepdfdatetime\pdfcreationdate\endparsepdfdatetime

  \settimeformat{ampmtime}
  \newcommand{\version}[1]{\emph{Version #1 (Built:~\today~@ \currenttime~UTC\thetimezone)}}
  \AtBeginShipout
  {\AtBeginShipoutAddToBox{
      \begin{tikzpicture}[overlay, remember picture]
      \node[anchor=north] at (current page.north) {\textcolor{blue}{\vspace{3em}\version{\versionnum}}};    
      \end{tikzpicture}
  }}
\fi

\newcommand{\hcfirst}[0]{$\mathrm{HC}_{\mathrm{first}}$}
\newcommand{\hcfirstbold}[0]{$\mathrm{\textbf{HC}}_{\mathrm{\textbf{first}}}$}

  \ifcameraready
\else
  
  \fancypagestyle{firststyle}
  {
    \fancyfoot[C]{\thepage}
  }
\fi

\newcites{S}{Survey Sources}

\begin{document}

\bstctlcite{IEEEexample:BSTcontrol}
\bstctlcite[@auxoutS]{IEEEexample:BSTcontrol}

\title{\mpi{A Case for Transparent Reliability in DRAM Systems}}

\newcommand{\ethz}{{\large$^\dagger$}} 
\newcommand{\tud}{{\large$^\ddagger$}}
\newcommand{\scomma}{{\large$^,$}}

\author{ \vspace{-2ex}\\%
Minesh Patel\ethz{}\hspace{0.15in}%
Taha Shahroodi\tud{}\ethz{}\hspace{0.15in}%
Aditya Manglik\ethz{}\hspace{0.15in}%
A. Giray Ya{\u{g}}l{\i}k{\c{c}}{\i}\ethz{}\hspace{0.15in}\\%
Ataberk Olgun\ethz{}\hspace{0.15in}%
Haocong Luo\ethz{}\hspace{0.15in}%
Onur Mutlu\ethz{}\vspace{2mm}\\%
\textit{\ethz{}ETH Z{\"u}rich\hspace{0.12in}\tud{}TU Delft}%
\\}

\maketitle

\newcommand\blfootnote[1]{%
  \begingroup
  \renewcommand\thefootnote{}\footnotetext{#1}%
  \addtocounter{footnote}{-1}%
  \endgroup
}

\begin{abstract}
  Mass-produced commodity DRAM is the preferred choice of main memory for a broad
range of computing systems due to its favorable cost-per-bit. However, today's
systems have diverse system-specific needs (e.g., performance, energy,
reliability) that are difficult to address using one-size-fits-all
general-purpose DRAM. Unfortunately, although system designers can theoretically
adapt commodity DRAM chips to meet their particular design goals (e.g., by
exploiting slack in access timings to improve performance, or implementing
system-level RowHammer mitigations), we observe that designers today lack the
necessary insight into commodity DRAM chips' reliability characteristics to
implement these techniques in practice. In this work, we \mpo{make a case} for
DRAM manufacturers to provide increased transparency into simple device
characteristics (e.g., internal row address mapping, cell array organization)
that affect consumer-visible reliability. Doing so has negligible impact on
manufacturers given that these characteristics can be reverse-engineered using
known techniques; however, \mpo{it} has significant benefit for system designers, who
can then make informed decisions to better adapt commodity DRAM to meet modern
systems' needs while preserving its cost advantages.

To support our argument, \mpo{we study four ways that system designers can adapt
commodity DRAM chips to system-specific design goals: (1) improving DRAM
reliability; (2) reducing DRAM refresh overheads; (3) reducing DRAM access
latency; and (4) defending against RowHammer attacks. We observe that adopting
solutions for any of the four goals requires system designers to make
assumptions about a DRAM chip's reliability characteristics. These assumptions
discourage system designers from using such solutions in practice due to the
difficulty of both making and relying upon the assumption.}

We identify DRAM standards as the root of the problem: current standards rigidly
enforce a fixed operating point \mpi{with no specifications for how a system
designer might explore alternative operating points.} \mpo{To overcome this
problem}, we introduce a two-step approach that reevaluates DRAM standards with
a focus on transparency of reliability characteristics so that system designers
are encouraged to make the most of commodity DRAM technology for both current
and future DRAM chips.
\end{abstract}  

\section{Introduction}
\label{sec:intro}

\mpm{Dynamic Random Access Memory (DRAM)~\cite{dennard1968field,
dennard1974design, keeth2007dram, markoff2019ibm, nature2018memory,
ibm2021dram}} is the dominant choice for main memory across a broad range of
computing systems because of its high capacity at low cost relative to other
viable main memory technologies. Building efficient DRAM chips requires
substantially different manufacturing processes relative to standard CMOS
fabrication~\cite{kim1999assessing}, so DRAM is typically designed and
manufactured separately from other system components. \mpt{In this way, system
designers who purchase, test, and/or integrate commodity DRAM chips (e.g., cloud
system designers, processor and system-on-a-chip (SoC) architects, memory module
designers, test and validation engineers) are free to focus on the particular
challenges of the systems they work on instead of dealing with the nuances of
building low-cost, high-performance DRAM.}

To ensure that system designers can integrate commodity DRAM chips from any
manufacturer, the DRAM interface and operating characteristics have long been
standardized by the JEDEC consortium~\cite{jedec2021jc42}. JEDEC maintains a
limited set of \emph{DRAM standards} for commodity DRAM chips with different
target applications, e.g., general-purpose DDR\emph{n}~\cite{jedec2008ddr3,
jedec2012ddr4, jedec2020ddr5}, \mpk{bandwidth-optimized
HBM\emph{n}~\cite{jedec2021high, jedec2022high},} mobile-oriented
LPDDR\emph{n}~\cite{jedec2014lpddr4, jedec2020lpddr5}, graphics-oriented
GDDR\emph{n}~\cite{jedec2016gddr5, jedec2016gddr6}. Given that DRAM designs are
heavily constrained by DRAM standards, manufacturers generally seek
profitability through economies of scale~\cite{kang2010study, dell1997white,
lee2013strategic, croswell2000model}: they mass produce standards-compliant DRAM
chips using highly-optimized \mpi{manufacturing} processes. High-volume
production amortizes manufacturing costs and increases per-chip profit margins.
As such, DRAM manufacturers conservatively regard design- and
manufacturing-related information as sensitive~\mpo{\cite{nair2013archshield,
gong2017dram, childers2015achieving, cost1997yield}, revealing only what DRAM
standards require.}

To maintain their competitive advantage in cost-per-capacity, DRAM manufacturers
continually improve storage densities across successive product generations
while minimizing fabrication costs (e.g., minimizing chip area, maximizing
yield). This requires a careful balance between aggressively scaling physical
feature sizes, continually optimizing circuit designs to reduce area
consumption, and mitigating reliability issues that arise with process
technology shrinkage~\mpo{\cite{kang2014co, cha2017defect, nair2013archshield,
micron2017whitepaper, park2015technology, son2015cidra, micron2022quaterly}}.
Unfortunately, focusing primarily on storage density forces DRAM manufacturers
to sacrifice potential improvements in other metrics of interest, such as
performance, energy, etc. Even if process technology shrinkage naturally
\mpo{provides} gains in these other metrics (e.g., by reducing circuit latencies
with smaller circuit elements), manufacturers typically adjust their designs to
exchange these gains for additional storage density (e.g., by building larger
array sizes that offset any reductions in access latency). As manufacturers
juggle the complex tradeoffs in chip design and manufacturing to maintain market
competitiveness, DRAM as a whole exhibits slow generational improvements in key
areas, such as access latency and power consumption~\mpo{\cite{chang2017thesis,
lee2016reducing, ghose2018your}}. 

\mpi{Figure~\ref{fig:timings_idds} provides a best-effort survey showing how
manufacturer-reported values for four key DRAM operating timings and per-chip
storage capacity \mpn{(all shown in log scale)} have evolved over time. We
extract these data values from 58 publicly-available DRAM chip datasheets from
across 19 different DRAM manufacturers with datasheet publication dates between
1970 and 2021. This data encompasses DRAM chips from both asynchronous (e.g.,
page mode, extended data out) and synchronous (e.g., SDRAM, DDR\emph{n}) DRAM
chips. Appendix~\ref{position:appendix_a} describes our data collection
methodology in further detail, and Appendix~\ref{position:appendix_b} provides
an overview of our dataset, which is publicly available on
GitHub~\cite{datasheetsurveygithub}.}

\noindent
\begin{minipage}{\linewidth}
\renewcommand{\thempfootnote}{\roman{mpfootnote}}
\begin{figure}[H]
    \centering
    \includegraphics[width=\linewidth]{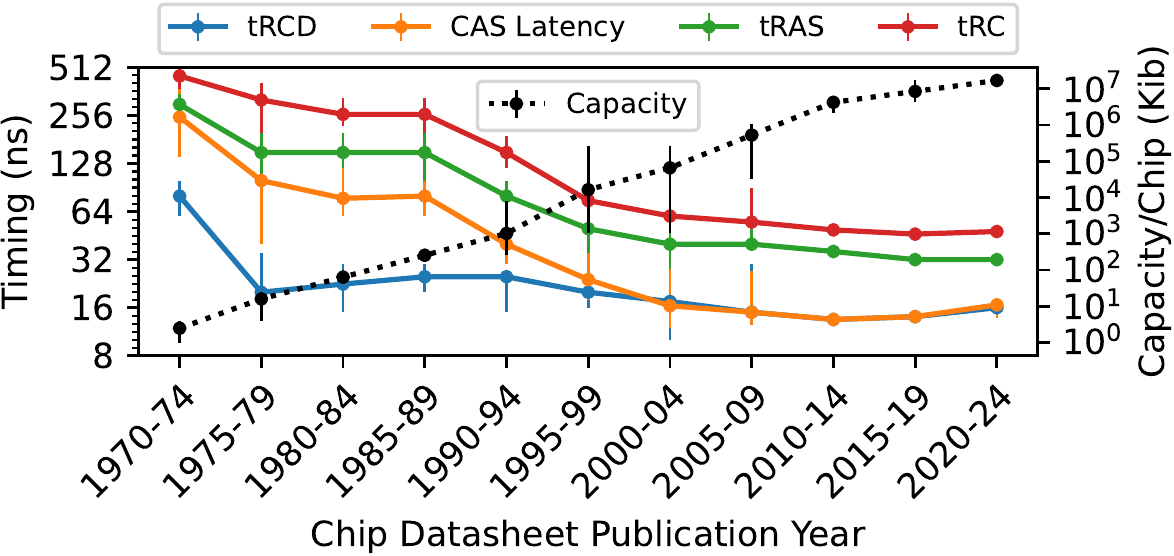}
    \caption[]{\mpo{%
    \mpn{Semi-log plot showing the evolution of key} DRAM access timings (left) and per-chip storage capacity (right)\protect\footnote{%
    \mpo{JEDEC-standardized parameters\cite{jedec2020ddr5} found in DRAM chip datasheets:}\\
    \indent\indent\indent
    \begingroup%
    \scriptsize
    \begin{tabular}{ll}%
        \textbf{Parameter} & \textbf{Definition}\\\hline
        tRCD & minimum row activation to column operation delay\\
        CAS Latency & read operation to data access latency\\
        tRAS & minimum row activation to precharge delay\\
        tRC & minimum delay between accesses to different rows
    \end{tabular}%
    \endgroup%
    }\, \mpn{across each 5-year period of time.}}}
    \label{fig:timings_idds}
\end{figure}
\end{minipage}

\mpi{We observe a clear trend that newer DRAM chips exhibit improvements in all
four timing parameters \emph{and} storage capacity. However, \emph{none} of the
four timings have improved significantly in the last two decades. For example,
the median tRCD/CAS Latency/tRAS/tRC reduced by 2.66/3.11/2.89/2.89\% per year
on average between 1970 and 2000, but only 0.81/0.97/1.33/1.53\% between 2000
and 2015.\footnote{We report 2015 instead of 2020 because 2020 shows a
regression in CAS latency due to first-generation DDR5 chips, which we believe
is not representative because of its immature technology.} In contrast, storage
capacity improved relatively consistently with an exponential growth factor of
0.328 per year (0.341 for 1970-2000 and 0.278 for 2000-2020) across the entire
history of DRAM technology. This data is consistent with similar \mpj{studies}
done in prior work~\mpm{\cite{son2013reducing, chang2017thesis, lee2013tiered,
hennessy2011computer, chang2016understanding, lee2016reducing,
isaac2008remarkable, choi2015multiple, borkar2011future,
nguyen2018nonblocking}}, showing that commodity DRAM manufacturers have
prioritized storage capacity over access latency in recent years.}

Unfortunately, prioritizing storage density does not always align with the
increasingly diverse needs of modern computing systems. These needs change as
systems continuously evolve, so there is no single target metric (e.g., storage
capacity) that suits all DRAM-based systems. Instead, each system's design goals
differ based on factors such as cost, complexity, applications, etc. For
example, storage-focused data centers (e.g., content delivery network nodes) may
require high-reliability memory while compute-focused clusters may optimize for
performance with low-latency memory. Unfortunately, system designers today are
limited to a narrow range of commodity DRAM products,\footnote{\mpi{Custom DRAM
chips (e.g., latency-optimized~\cite{micron2021rldram, fujitsu2012fcram},
high-reliability~\cite{smart2021rugged, im2016im}) and target-specific chips
(e.g., LPDDR\emph{n}~\cite{jedec2014lpddr4, jedec2020lpddr5},
GDDR\emph{n}~\cite{jedec2016gddr5, jedec2016gddr6}) sacrifice the cost
advantages of high-volume general-purpose commodity
DRAM~\cite{lee2013strategic}.}} that effectively restrict design freedom and
limit the peak potential of DRAM-based systems. 

To address this disparity, system designers have long since developed techniques
for adapting unmodified commodity DRAM chips to varying system requirements.
Examples include: \mpl{(1)} actively \mpi{identifying and/or} mitigating errors
to improve reliability~\mpi{\cite{kim2015bamboo, cardarilli2000development,
yoon2010virtualized, udipi2012lot, jian2013low, kim2015frugal, nair2016xed,
jian2013adaptive, han2014data, chen2015ecc, chen2013e3cc, manzhosov2021muse,
patil2021dve, choi2020reducing, sharifi2017online, alameldeen2011energy,
naeimi2013sttram, awasthi2012efficient}}; \mpl{(2)} exploiting available
timing~\mpi{\cite{chandrasekar2014exploiting, chang2016understanding,
kim2018solar, lee2015adaptive, lee2017design, wang2018reducing,
zhang2016restore, koppula2019eden}} and
voltage~\mpl{\cite{chang2017understanding, david2011memory, deng2011memscale}}
margins to reduce memory access latency, power consumption, decrease refresh
overheads~\cite{venkatesan2006retention, liu2012raidr, nair2013archshield,
patel2017reach, lin2012secret, ghosh2007smart, wang2018content,
qureshi2015avatar, patel2017reach}; and \mpl{(3)} mitigating unwanted DRAM
\mpo{data persistence~\cite{halderman2008lest, gruhn2013practicability,
simmons2011security} and \mpo{read-disturb}
problems~\cite{yaglikci2020blockhammer, kim2014flipping, apple2015about,
kim2021mithril, saileshwar2022randomized}.}
Section~\ref{subsec:adaptability_benefits_consumers} discusses these proposals
in greater detail to motivate the need to adapt commodity DRAM to diverse yet
aggressive design targets.

However, these proposals are largely theoretical ideas or proofs-of-concept
based on performance and reliability characteristics that are
\mpo{\emph{assumed}, \emph{inferred}, or \emph{reverse-engineered} from a
limited set of observations and DRAM products (e.g., in-house experimental
studies)} without DRAM manufacturers' support. \mpo{Therefore, adopting such
proposals in a consumer-facing product requires a system designer to weigh the
benefits of enhancing DRAM (e.g., improving performance, security, etc.) against
both: (1) risks (e.g., failures in the field) associated with potentially
violating manufacturer-recommended operating conditions and (2) limitations due
to compatibility with only a subset of all commodity DRAM products (e.g., only
those that have been accurately reverse-engineered). These risks and limitations
are a serious barrier to adoption, especially for small-scale designers who may
have limited headroom and expertise} for exploring \mpt{unconventional designs}.

In this work, we argue that \emph{the lack of transparency} concerning DRAM
reliability characteristics is ultimately responsible for confining system
designers to conventional, specification-constrained designs. For example,
safely improving DRAM access latency by adjusting operating timings requires
understanding the possible failure modes resulting from using non-standard
timings (discussed further in Section~\ref{sec:latency_study}). This is because
selecting suitable operating timings requires the system designer to estimate
the reliability impact of the new timings, which in turn requires reliability
modeling or extensive testing under worst-case operating conditions.
\mpo{Unfortunately, obtaining the information necessary to make these estimates
(e.g., error models, worst-case testing parameters) is difficult, if not
impossible,\footnote{\mpo{For all but the largest customers capable of
independently conducting rigorous post-manufacturing testing.}} without
transparency from DRAM manufacturers. This transparency does not exist today,
even through private agreements for high-volume consumers who have significant
stake in the DRAM industry~\mpk{\cite{safari2016private, saroiu2022price,
loughlin2021stop}}.} In general, without the ability to understand how different
design choices can impact DRAM reliability (e.g., error rates), system designers
are discouraged from \mpo{deploying or even} exploring alternative designs.

To understand the source of the transparency problem, we conduct four case
studies \mpi{throughout Sections~\ref{sec:rela_study}--\ref{sec:security_study}}
that each examine a key system design concern for commodity DRAM chips: \mpi{(1)
reliability; (2) refresh overheads; (3) access latency; and (4) the RowHammer
security vulnerability. For each case study, we explain how system designers are
forced to make assumptions about DRAM reliability in order to address these
concerns without breaking design independence with DRAM manufacturers, but those
very assumptions limit the practicality and scope of the solution.} We then
argue that DRAM standards lie at the heart of the problem because they do not
adequately address the aforementioned DRAM reliability concerns. \mpo{To
overcome this reliance on assumptions, we show that incorporating specifications
for consumer-visible DRAM reliability characteristics (e.g., industry-validated
error models and testing techniques) into DRAM standards} alleviates the problem
and allows system designers to better adapt commodity DRAM to their particular
needs without requiring \mpo{changes to} how DRAM manufacturers design and build
commodity DRAM.

\mpt{We propose incorporating information transparency into DRAM standards using
a two-step approach involving all DRAM stakeholders, including consumers and
manufacturers. In Step 1, for DRAM chips already in the field, we seek the
release of basic information} about DRAM chips that consumers can use to better
understand the chips' reliability characteristics.
Section~\ref{position:subsec:what_to_release} details examples of possible
information to release, including (1) basic microarchitectural characteristics
(e.g., organization of physical rows, sizes of internal storage arrays) that can
be reverse-engineered using existing techniques \mpi{with access to appropriate
testing infrastructure}~\mpm{\cite{jung2016reverse, lee2015adaptive,
patel2019understanding, chang2016understanding, mukhanov2020dstress,
kim2018dram, kraft2018improving, kim2018solar, orosa2021deeper,
hassan2021uncovering}} and (2) industry-recommended testing best practices
(e.g., test patterns for key error mechanisms). \mpt{We believe that this
information can be released through a combination of (1) crowdsourced testing of
commodity DRAM chips on the market; and (2) DRAM chip manufacturers publishing
information (e.g., using datasheet revisions or online resources) about their
products, possibly limited to basic information that manufacturers already have
available (i.e., that requires minimal logistical effort to release). Through a
combination of these two avenues,} \mpo{\mpi{information can be provided to
\emph{all} system designers, including the majority of designers without the
ability to conduct exhaustive testing,} almost immediately without requiring
changes to existing DRAM hardware or standards (though standardizing the
information release could streamline the process). Then, armed with this
information, system designers can make more informed decisions when developing
their own solutions to system-specific design concerns while also preserving the
advantages of commodity DRAM built per general-purpose DRAM standards.}

\mpo{In Step 2, we propose extending DRAM standards with explicit DRAM
reliability standards that provide industry-standard guarantees, tools, and/or
information helpful to consumers. We envision different possibilities for these
reliability standards, including (1) reliability guarantees for how a chip is
expected to behave under certain operating conditions (e.g., predictable
behavior of faults~\cite{criss2020improving}); (2) disclosure of
industry-validated DRAM reliability models and testing strategies suitable for
commodity DRAM chips (e.g., similar to how JEDEC JEP122~\cite{jedec2016failure},
JESD218~\cite{jedec2010ssdrequirements}, and
JESD219~\cite{jedec2010ssdendurance} address \mpo{Flash-memory-specific error
mechanisms~\mpm{\cite{cai2017error, cai2012error, cai2018errors}} such as
floating-gate data retention~\mpm{\cite{cai2015data, luo2018heatwatch,
luo2018improving, cai2012flash} and \mpm{models for physical phenomena such as
threshold voltage distributions~\cite{cai2013threshold, cai2013program,
cai2015read, luo2016enabling}}}}); and (3) requirements for manufacturers to
directly provide relevant information about their DRAM chips (e.g., the
information requested in Step 1).} As the DRAM industry continues to evolve, we
anticipate closer collaboration between DRAM and system designers to efficiently
overcome the technology scaling challenges that DRAM is already
facing~\mpm{\cite{kang2014co, micron2017whitepaper, mutlu2013memory,
mutlu2015main}}. Although we hope that transparency will occur naturally as part
of this process, we believe the end result will be determined in a large part by
the direction in which DRAM standards evolve. Therefore, we believe that
ensuring transparency of reliability characteristics becomes a first-order
concern is essential for allowing innovation going forward.

\mpo{We make the following contributions:}

\begin{enumerate}
    \item \mpo{We make a case for the DRAM industry to provide transparency into
    the consumer-visible reliability characteristics of commodity DRAM chips so
    that system designers can make informed decisions when integrating commodity
    chips into their designs.}

    \item \mpo{We support our argument with four case studies (DRAM reliability,
    DRAM refresh, DRAM access latency, and RowHammer), showing that system
    designers require insight into commodity DRAM chip reliability in order to
    adopt improvements in any of the four directions.}

    \item \mpo{We identify modern DRAM standards as the primary factor that
    limits system designers from comprehensively understanding the reliability
    impact of their design decisions, thereby discouraging the designers from
    adopting techniques to better adapt commodity DRAM chips to their systems'
    specific needs.}
    
    \item \mpo{\mpt{We propose a new two-step approach to facilitate
    transparency into consumer-visible DRAM reliability characteristics. In the
    short term, we ask for information release through a combination of both (1)
    crowdsourced testing from DRAM consumers; and (2) official information from
    DRAM manufacturers, possibly standardized by extensions to DRAM
    standards.} In the long term, we propose extending DRAM standards
    with explicit DRAM reliability standards that provide industry-standard
    guarantees, tools, and/or information that enable DRAM consumers to perform
    their own reliability analyses and understand DRAM reliability at different
    operating points.}
\end{enumerate} 
\section{\xmp{The System Designer's Challenge}}
\label{sec:motivation_new}

Today's DRAM industry thrives on separation of concerns: DRAM manufacturers can
focus on designing highly-optimized DRAM chips while consumers can make use of
standardized DRAM \xmg{that conform to JEDEC standards}. This design
independence \xmt{is powerful because it allows each party to leverage their
respective expertise to build the best possible product. As a result, a system
designer who is responsible for choosing the memory substrate for a particular
system can simply select between a limited range of standardized commodity parts
that are optimized for different targets, such as general-purpose performance
(e.g., DDR\emph{n}~\cite{jedec2012ddr4, jedec2020ddr5}), high bandwidth (e.g.,
GDDR\emph{n}~\cite{jedec2016gddr5, jedec2016gddr6},
\mpo{HBM\emph{n}~\cite{jedec2021high, jedec2022high}}), and low power (e.g.,
LPDDR\emph{n}~\cite{jedec2014lpddr4, jedec2020lpddr5}).}

Unfortunately, the system designer faces a significant challenge: the designer
is unable to fully explore the memory design space \mpk{(as well as the
system-memory co-design space)} because there are only a limited number of
viable design points using commodity DRAM chips. Therefore, the limited number
of options inherently forces the designer to overlook opportunities for
customizing DRAM operation towards their system's particular design goals. As
main memory becomes an increasingly significant system
bottleneck~\mpm{\cite{mutlu2013memory, mutlu2014research, mutlu2019processing}},
we believe that enabling system designers to flexibly adapt commodity DRAM to
suit their own needs as they see fit is a promising path to reap the benefits of
adaptability while preserving the design independence between DRAM manufacturers
and system designers.

\subsection{\mpi{Benefits for DRAM Consumers}}
\label{subsec:adaptability_benefits_consumers}

Prior works~\mpk{\cite{kim2015bamboo, cardarilli2000development,
yoon2010virtualized, udipi2012lot, jian2013low, kim2015frugal, nair2016xed,
jian2013adaptive, chen2015ecc, chen2013e3cc, manzhosov2021muse, patil2021dve,
wang2018content, mcelog2021bad, nvidia2020dynamic, venkatesan2006retention,
baek2014refresh, hwang2012cosmic, meza2015revisiting, liu2012raidr,
ohsawa1998optimizing, wang2014proactivedram, lin2012secret, nair2013archshield,
ghosh2007smart, qureshi2015avatar, khan2014efficacy, khan2016case,
khan2016parbor, khan2017detecting, jafri2020refresh, kim2000dynamic,
kim2003block, katayama1999fault, patel2017reach, mathew2017using,
chandrasekar2014exploiting, chang2016understanding, kim2018solar,
lee2015adaptive, lee2017design, wang2018reducing, zhang2016restore,
hassan2016chargecache, koppula2019eden, mathew2017using, zhang2021quantifying,
gao2019computedram, kim2018dram, olgun2021pidram, olgun2021quac, kim2019d,
seshadri2013rowclone, seshadri2015fast, seshadri2017ambit, seshadri2019dram,
hajinazar2021simdram, seshadri2016buddy, yaglikci2020blockhammer,
greenfield2012throttling, mutlu2018rowhammer, kim2014flipping, konoth2018zebram,
van2018guardion, brasser2017can, saileshwar2022randomized}} demonstrate
significant system-level benefits from adapting commodity DRAM operation to
different system needs without changing the DRAM design itself. This section
reviews the benefits of four concrete examples of such customizations: \mpk{1)}
DRAM reliability improvement, \mpk{2)} DRAM refresh overhead reduction, \mpk{3)}
DRAM access latency reduction, and \mpk{4)} RowHammer security improvement. In
principle, a system designer can readily implement each customization using
existing techniques. Unfortunately, adopting these techniques in practice
requires understanding how DRAM reliability characteristics behave under
different operating conditions, which is not clearly communicated by DRAM
manufacturers or standards today. In this section, we review each example
\mpk{customization}'s potential benefits; then, our case studies throughout
Sections~\ref{sec:rela_study}-\ref{sec:security_study} explore each example
\mpk{customization} in further detail to identify the specific factors that we
believe discourage system designers from adopting the examples in practice.

\subsubsection{\mpo{DRAM Reliability Improvement}}
\label{subsubsec:mot_rela}

\mpo{DRAM is susceptible to a wide variety of error mechanisms that can impact
overall system reliability. To combat DRAM-related failures, system designers
typically incorporate reliability, availability and serviceability (RAS)
features~\cite{synopsys2015whitepaper, dell2008system, slayman2006impact} that
collectively improve system reliability beyond \mpk{what commodity DRAM chips
can provide alone}. In general, memory RAS is a broad research area with
solutions spanning the hardware-software stack, ranging from hardware-based
mechanisms within the \mpk{DRAM chip (e.g., on-die ECC
scrubbing~\cite{jedec2020ddr5, rahman2021utilizing, criss2020improving},
post-package repair~\cite{horiguchi2011nanoscale, jedec2012ddr4, jedec2020ddr5,
kim2016ecc, wada2004post}, target row refresh~\cite{hassan2021uncovering,
frigo2020trrespass})}, memory controller (e.g., rank-level
ECC~\cite{kim2015bamboo, cardarilli2000development, yoon2010virtualized,
udipi2012lot, jian2013low, kim2015frugal, nair2016xed, jian2013adaptive,
chen2015ecc, chen2013e3cc, manzhosov2021muse, patil2021dve, wang2018content},
rank-level ECC scrubbing~\mpk{\cite{han2014data, qureshi2015avatar,
choi2020reducing, sharifi2017online, alameldeen2011energy, naeimi2013sttram,
awasthi2012efficient, sridharan2015memory, rahman2021utilizing,
sharifi2017online}}, repair techniques~\cite{lin2012secret, nair2013archshield,
kline2020flower, longofono2021predicting, kline2017sustainable,
schechter2010use, nair2019sudoku, zhang2017dynamic, wang2017architecting,
kim2016relaxfault}) to software-only solutions (e.g., page
retirement~\cite{mcelog2021bad, nvidia2020dynamic, venkatesan2006retention,
baek2014refresh, hwang2012cosmic, meza2015revisiting}, failure
prediction~\cite{mukhanov2019workload, baseman2016improving,
giurgiu2017predicting, lan2010study, liang2006bluegene, boixaderas2020cost}).}

\mpo{As a specific and relevant example, an important category of hardware-based
redundancy mechanisms known as rank-level error-correcting codes (rank-level
ECC) operate within the memory controller to isolate the rest of the system from
random DRAM errors. Depending on the ECC design, rank-level ECC can protect
against random single-bit (e.g., SEC/SEC-DED Hamming
codes~\cite{hamming1950error}), multi-bit (e.g., BCH~\cite{hocquenghem1959codes,
bose1960class}, Reed-Solomon~\cite{reed1960polynomial}), and/or multi-component
(e.g., Chipkill~\cite{dell1997white, kim2015bamboo}) errors with varying
hardware and runtime overheads. The system designer must decide which ECC
mechanism is most appropriate for their particular system (e.g., which error
mechanisms are dominant and what degree of protection is required). For example,
a state-of-the-art rank-level ECC mechanism called}
\mpi{Frugal-ECC~\cite{kim2015frugal} uses data compression to provide
chipkill-correct ECC for $\times$4 non-ECC DIMMs and $\times$8 ECC DIMMs with
negligible performance (maximum of 3.8\%), energy-efficiency, and area overheads
compared with an industry-standard chipkill solution. Therefore, Frugal-ECC
enables system designers to implement chipkill reliability using commodity DRAM
chips with a fraction of the storage overheads suffered by conventional ECC DIMM
configurations.}

\subsubsection{DRAM Refresh Overhead Reduction}
\label{subsubsec:mot_dram_refresh}

DRAM stores data in volatile capacitors, which are susceptible to charge
leakage. To prevent this leakage from causing data loss, DRAM requires periodic
refresh operations that intermittently access all DRAM cells to restore their
charge levels to safe values. Unfortunately, DRAM refresh operations are well
known to waste significant system performance and
power~\mpm{\cite{ohsawa1998optimizing, kim2000dynamic, laudon2006ultrasparc,
liu2012raidr, nair2013archshield, bhati2015flexible, wang2014proactivedram,
baek2014refresh, mathew2017using, bhati2016dram}}, sacrificing almost half of
the total memory throughput and wasting \mpk{almost} half of the total DRAM
power for projected 64 Gb chips~\cite{liu2012raidr}.

To alleviate the power and performance costs of DRAM refresh, prior
works~\mpo{\cite{liu2012raidr, ohsawa1998optimizing, wang2014proactivedram,
venkatesan2006retention, lin2012secret, nair2013archshield, ghosh2007smart,
qureshi2015avatar, khan2014efficacy, khan2016case, khan2016parbor,
khan2017detecting, jafri2020refresh, kim2000dynamic, kim2003block,
katayama1999fault, patel2017reach, mathew2017using}} take advantage of the fact
that \emph{most} refresh operations are unnecessary.\footnote{Latency-hiding
techniques (e.g, prefetching, memory command scheduling, on-chip caching, etc.)
and parallelization of refresh and access
operations~\mpk{\cite{chang2014improving, nguyen2018nonblocking, pan2019hiding,
stuecheli2010elastic, mukundan2013understanding}} help mitigate performance
overheads but do not change the total number of refresh operations issued.
\mpk{As a result, such techniques cannot mitigate energy wastage due to DRAM
refresh}. These techniques are also imperfect in many cases \mpk{where
latency-hiding is impractical (e.g., row conflicts between refresh and access
commands, larger memory footprints than available caching
resources)~\mpm{\cite{nair2013case, pan2019hiding, zhang2014cream,
chang2014improving}}.}} The standard DRAM refresh algorithm refreshes all cells
frequently (i.e., at the worst-case rate) to simplify DRAM refresh and guarantee
correctness. However, each cell's data retention characteristics vary
significantly due to a combination of data-dependence~\mpk{\cite{khan2016parbor,
khan2017detecting, liu2013experimental, khan2014efficacy, patel2017reach}} and
process variation~\mpo{\cite{hamamoto1995well, hamamoto1998retention,
gong2017dram, nair2013archshield, liu2013experimental, liu2012raidr,
wang2014proactivedram}}. \mpi{As a result, eliminating unnecessary refresh
operations} can provide significant \mpk{power reduction and performance
improvement.} For example, Liu et al.~\cite{liu2012raidr} demonstrate an average
\mpl{energy-per-access and system performance improvement of 8.3\% and 4.1\%,
respectively, for 4~Gib chips (49.7\% and 107.9\% for 64~Gib chips)} when
relaxing the refresh rate at the row granularity. Therefore, reducing refresh
overheads can potentially benefit any DRAM-based system.

\subsubsection{\xmt{DRAM Access Latency Reduction}}
\label{subsubsec:mot_dram_latency}

Figure~\ref{fig:timings_idds} shows that DRAM access latency has not
significantly improved relative to storage capacity over the last two decades.
This makes DRAM an increasingly significant system performance bottleneck today,
especially for workloads with large footprints that are sensitive to DRAM access
latency~\cite{hsieh2016accelerating, ferdman2012clearing, gutierrez2011full,
hestness2014comparative, huang2014moby, zhu2015microarchitectural,
oliveira2021damov, boroumand2018google, boroumand2021google, koppula2019eden,
kanellopoulos2019smash, son2013reducing, mutlu2013memory, wilkes2001memory,
wulf1995hitting, mutlu2007stall, mutlu2003runahead, kanev2015profiling,
mutlu2014research, bera2019dspatch, bera2021pythia, liu2019binary,
ghose2019processing, shin2014nuat, ghose2019demystifying}. Although
\mpk{conventional} latency-hiding techniques (e.g., caching, prefetching,
multithreading) can potentially help mitigate many of the performance concerns,
these techniques (1) fundamentally do not change the latency of each memory
access and (2) fail to work in many cases (e.g., irregular memory access
patterns, random accesses, huge memory footprints).

To address this problem, prior works have taken two major directions. First,
many works~\mpk{\cite{chandrasekar2014exploiting, chang2016understanding,
kim2018solar, lee2015adaptive, lee2017design, wang2018reducing,
zhang2016restore, hassan2016chargecache, koppula2019eden, mathew2017using,
zhang2021quantifying}} show that the average DRAM access latency can be
shortened by reducing DRAM access timings for particular memory locations that
can tolerate faster accesses. This can be done safely because, although DRAM
standards call for constant access timings across all memory locations, the
minimum viable access timings that the hardware can support actually differ
between memory locations due to factors such as heterogeneity in the circuit
design~\mpo{\cite{lee2017design, lee2016reducing}} and manufacturing process
variation between circuit components~\mpm{\cite{chandrasekar2014exploiting,
chang2016understanding, kim2018solar, chang2017understanding, lee2015adaptive}}.

\xmt{Exploiting these variations in access timings to reduce the average memory
access latency can provide significant system performance improvement. For
example, Chang et al.~\cite{chang2016understanding} experimentally show that
exploiting access latency variations can provide an average 8-core system
performance improvement of 13.3\%/17.6\%/19.5\% for real DRAM chips from three
major DRAM manufacturers. Similarly, Kim et al.~\cite{kim2018solar} show that
exploiting access latency variations induced by DRAM sense amplifiers provides
an average (maximum) system performance improvement of 4.97\% (8.79\%) versus
using default DRAM access timings for 4-core heterogeneous workload mixes
\mpo{based on data obtained from 282 commodity LPDDR4 DRAM chips}.}

Second, other works~\mpk{\cite{gao2019computedram, kim2018dram, olgun2021pidram,
olgun2021quac, kim2019d, seshadri2013rowclone, seshadri2015fast,
seshadri2017ambit, seshadri2019dram, hajinazar2021simdram, seshadri2016buddy}}
show that commodity DRAM can perform massively-parallel computations (e.g., at
the granularity of an 8 KiB DRAM row) by exploiting \mpk{the underlying analog
behavior of DRAM operations} (e.g., charge sharing between cells). These works
show that such computations can significantly improve overall system performance
and energy-efficiency by both (1) reducing the amount of data transferred
between the processor and DRAM and (2) exploiting the relatively high throughput
of row-granularity operations. For example, Gao et al.~\cite{gao2019computedram}
show that in-DRAM 8-bit vector addition is $9.3\times$ more energy-efficient
than the same computation in the processor, primarily due to avoiding the need
for off-chip data transfers. \mpo{Similarly, Olgun et al.~\cite{olgun2021pidram}
use an end-to-end FPGA-based evaluation infrastructure to demonstrate that
in-DRAM copy and initialization techniques can improve the performance of
system-level copy and initialization by $12.6\times$ and $14.6\times$,
respectively.}

\subsubsection{\mpo{Improving Security Against RowHammer}}
\label{subsubsec:mot_dram_security}

RowHammer~\mpk{\cite{kim2014flipping, bains2014row, mutlu2017rowhammer,
mutlu2019rowhammer}} is a well-studied read-disturb phenomenon in modern DRAM
chips in which memory accesses to a given memory location can induce bit-flips
at other locations. Recent experimental studies~\cite{kim2014flipping,
kim2020revisiting} show that RowHammer is continually worsening with process
technology shrinkage. Although DRAM manufacturers incorporate internal
RowHammer-mitigation mechanisms~\cite{de2021smash, lee2014green,
frigo2020trrespass, hassan2021uncovering, cojocar2019exploiting,
kim2020revisiting, micron20208gb}, prior work~\mpk{\cite{frigo2020trrespass,
cojocar2020are, hassan2021uncovering, jattke2022blacksmith, de2021smash}} shows
that these mechanisms do not suffice. Therefore, several
works~\mpo{\cite{yaglikci2020blockhammer, park2020graphene,
kim2014flipping, yaglikci2021security, aichinger2015ddr, apple2015about}}
provide RowHammer-mitigation mechanisms that operate from outside of the DRAM
chip to provide strong security without requiring changes \mpo{to DRAM chip
hardware} or relying upon \mpo{information} from DRAM manufacturers. Such a
solution is attractive for a system designer with interest in building a secure
system because the designer can rely upon their own methods rather than relying
upon external, possibly difficult-to-verify \mpo{promises or}
guarantees~\mpk{\cite{saroiu2022price, qureshi2021rethinking}}.

\mpf{Following prior work~\cite{yaglikci2020blockhammer}, we classify
previously-proposed RowHammer defenses into four different categories as
follows.}
\begin{enumerate}
    \item \emph{Access-agnostic} mitigation hardens a DRAM chip against
    RowHammer independently of the memory access pattern. \mpk{This includes
    increasing the overall DRAM refresh rate~\cite{kim2014flipping,
    apple2015about, aichinger2015ddr} and memory-wide error correction and/or
    integrity-checking mechanisms such as strong
    ECC~\cite{qureshi2021rethinking, cojocar2019exploiting, kim2014flipping}.
    These mechanisms are algorithmically simple but can introduce significant
    system hardware, performance, and/or energy-efficiency overheads (e.g.,
    \mpm{a large number of} additional refresh operations~\cite{kim2014flipping,
    kim2020revisiting, bhati2016dram}).} 
    
    \item \mpf{\emph{Proactive} mitigations~\cite{yaglikci2020blockhammer,
    greenfield2012throttling, mutlu2018rowhammer, kim2014flipping} adjust the
    DRAM access pattern to prevent the possibility of RowHammer errors.}
    
    \item \mpf{\emph{Physically isolating} mitigations~\cite{konoth2018zebram,
    van2018guardion, brasser2017can, saileshwar2022randomized, hassan2019crow}
    physically separate data such that accesses to one portion of the data
    cannot cause RowHammer errors in another.}
    
    \item \mpf{\emph{Reactive} mitigations~\cite{kim2014flipping,
    aweke2016anvil, son2017making, seyedzadeh2018cbt, you2019mrloc,
    lee2019twice, park2020graphene, kim2014architectural, kang2020cattwo,
    bains2015row, jedec2020ddr5, bains2016distributed, bains2016row,
    devaux2021method, yaglikci2021security, marazzi2022protrr,
    kim2015architectural} identify symptoms of an ongoing RowHammer attack
    (e.g., excessive row activations) and issue additional row \mpl{activation
    or} refresh operations to prevent bit-flips from occurring.}
\end{enumerate}

\noindent
RowHammer defense is an ongoing area of research, and which mechanism type is
most effective depends on the level of security (e.g., the threat model) that
the system designer requires and the trade-offs \mpk{(e.g., performance, energy,
hardware area, complexity overheads)} they are willing to make.

\begin{figure*}[b]
    \centering
    \includegraphics[width=\linewidth]{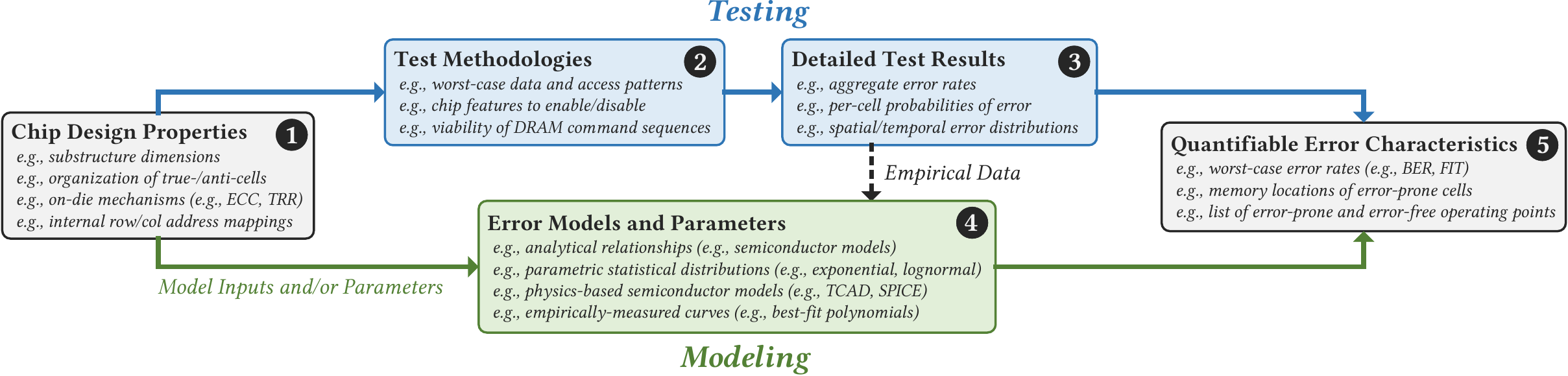}
    \caption{\mpk{Flow} of information necessary to determine key error
    characteristics for a given DRAM device.}
    \label{fig:test_flow}
\end{figure*}

\subsection{\mpi{Benefits for DRAM Manufacturers}}
\label{subsec:adaptability_benefits_manufacturers}

We believe that the ability to adapt commodity DRAM to system-specific
\mpk{design goals} also benefits DRAM manufacturers for two key reasons. First,
adaptability broadens the scope and competitive advantage of DRAM technology
relative to alternative technologies (e.g., emerging memories). Second, enabling
DRAM consumers to more easily innovate on the DRAM substrate can encourage
valuable feedback for DRAM manufacturers, including insights from customer
use-cases and \mpk{well-evaluated} suggestions for future products.

\mpi{Regardless of these benefits,} we believe making commodity DRAM adaptable
\mpi{has no significant downside for} DRAM manufacturers. The reliability
characteristics that we wish to be communicated \mpk{(as described in detail in
Section~\ref{position:subsec:what_to_release})} are either (1) already exposed
in scientific studies today; or (2) can be reverse-engineered using existing
techniques by those with access to appropriate tools (e.g., competitors,
scientific labs). We simply ask for these characteristics to be officially
provided in a trustworthy capacity. DRAM manufacturers have not previously
provided this information because there has been no pressing need to do so.
However, releasing this information makes sense today because it can enable a
broad range of benefits for DRAM consumers going forward, especially as DRAM
technology scaling continues to face increasing
difficulties~\mpk{\cite{mutlu2013memory, mutlu2021primer, mutlu2017rowhammer}}.

\subsection{\xmp{Short-Term vs. Long-Term Solutions}}
\label{subsec:mot_short_vs_long_term}

Prior works~\mpm{\cite{patterson1997case, mutlu2014research, mutlu2021primer,
mutlu2013memory, kim2014flipping, kang2014co, mutlu2017rowhammer,
mutlu2019rowhammer}} have praised the merits of cooperation between DRAM
manufacturers and system designers in order to collaboratively solve main memory
challenges across the system stack. However, this requires either (1) breaking
design independence between the two parties; (2) achieving consensus among all
DRAM stakeholders (i.e., JEDEC committee members and representatives, including
DRAM manufacturers and consumers) for every design change, followed by a lengthy
adoption period; \mpk{or (3) reducing dependence on DRAM standards and JEDEC. We
do not believe any of these options are easy to adopt} for either the (1) short
term, where we would like to quickly effect changes that enable information
transparency; or (2) long term, where breaking design independence constrains
the very freedom that we advocate system designers should have in meeting their
own design goals while preserving the cost advantages of mass-produced commodity
DRAM chips.

Instead, we argue for enabling each party to solve their own system-specific
design challenges, modifying DRAM standards only for issues that
collectively affect all DRAM stakeholders. However, regardless of how the DRAM
industry evolves over the coming years, we firmly believe that DRAM must become
more adaptable, whether that occurs through standards or
collaboration. 
\section{\mpk{Quantitatively Measuring Reliability}}
\label{position:sec:formalizing}

As we will show in the following case studies
(Sections~\ref{sec:rela_study}--\ref{sec:security_study}), a system designer
exploring unconventional DRAM operating points must first understand how
reliably a chip will behave at that operating point. Given that this behavior is
not governed by DRAM standards or described by DRAM manufacturers, the system
designer must determine it themselves, e.g., through modeling and/or testing.
This section formalizes \mpk{the information that a system designer may need
(but does not necessarily have access to today) in order to quantitatively
understand DRAM reliability.} 

\subsection{Information Flow During Testing}

Figure~\ref{fig:test_flow} describes the flow of information necessary for a
\mpl{system designer to quantitatively estimate\footnote{``Estimate'' because,
in general, no \mpl{model or experiment} is likely to be perfect, including
those provided by manufacturers.} a DRAM chip's error characteristics
\circled{5} starting from basic properties of the chip \circled{1}. In
principle, these characteristics can comprise \emph{any} aspect of DRAM
reliability that a system designer wants to quantify while exploring their
system's design and/or configuration space. Examples include: (1) worst-case
error rates (e.g., bit error rate (BER) or failures in time (FIT)) across a
given set of operating points; (2) a profile of error-prone memory locations};
or (3) a list of error-free operating points \mpk{(e.g., as identified in a
shmoo analysis~\cite{baker1997shmoo})}. \mpl{The error characteristics can be
estimated in two different ways: testing or modeling.}

\subsubsection{\mpl{Determination from Testing}}

\mpl{First, a system designer may estimate error characteristics using
measurements from detailed experimental testing \circled{3} across a variety of
operating conditions. Examples of measured quantities include:} aggregate error
rates, per-cell probabilities of error, and spatial/temporal error
distributions. \mpk{These measurements can be made using testing infrastructures
ranging from industry-standard large-scale testing
equipment~\cite{advantest2022t5833, teradyne2022magnum} to home-grown tools
based on commodity FPGAs~\cite{hassan2017softmc, olgun2021pidram, hou2013fpga,
kim2014flipping, gao2019computedram, chang2017understanding, ghose2018your,
weis2015retention, chang2016understanding, khan2014efficacy, wang2018dram,
ladbury2013use} or DRAM-based computing systems~\mpl{\cite{passmark2019memtest,
cojocar2020are, veen2016drammer, francis2018raspberry, david2011memory}}.}

\mpl{To conduct accurate and rigorous testing, the system designer must use an
effective test methodology \circled{2} that suits the particular DRAM chip under
test. Prior works extensively study key aspects of effective test methodologies,
including} \mpt{appropriate data and access patterns, the effects of
enabling/disabling DRAM chip features such as \mpk{target row refresh
(TRR)~\cite{frigo2020trrespass, marazzi2022protrr, hassan2021uncovering,
jattke2022blacksmith, kim2020revisiting} and on-die error correcting codes
(on-die ECC)~\mpm{\cite{nair2016xed, micron2017whitepaper, kang2014co, oh20153,
gong2017dram, son2015cidra, oh2014a, kwak2017a, kwon2014understanding,
patel2019understanding, patel2020bit, patel2021enabling}}}, and the viability of
different DRAM command sequences (e.g., sequences that enable in-DRAM row copy
operations~\cite{seshadri2013rowclone, gao2019computedram, olgun2021pidram,
chang2016low}, \mpm{true} random-number generation~\cite{olgun2021quac,
kim2019d, talukder2018exploiting, bostanci2022dr}, and physically unclonable
functions~\cite{kim2018dram, talukder2018ldpuf}).} 

\mpl{In turn, choosing an effective test methodology requires knowledge of
basic} \mpt{properties about a DRAM chip's design and/or error mechanisms
\circled{1}. For example, DRAM manufacturer's design choices for the sizes of
internal storage arrays (i.e., mats~\mpm{\cite{lee2017design, zhang2014half,
son2013reducing, olgun2021quac}}), charge encoding conventions of each cell
(i.e., the true- and anti-cell organization~\cite{kraft2018improving,
liu2013experimental}), use of on-die reliability-improving mechanisms
\mpk{(e.g., on-die ECC, TRR)}, and organization of row and column addresses all
play key roles in determining if and how susceptible a DRAM chip is to key error
mechanisms (e.g., data retention~\cite{hamamoto1998retention,
kraft2018improving, patel2019understanding, liu2013experimental,
bacchini2014characterization, weber2005data, yamaguchi2000theoretical},
access-latency-related failures~\cite{kim2018solar, lee2013tiered,
lee2015adaptive, lee2017design, chang2016understanding, olgun2021quac,
koppula2019eden, chandrasekar2014exploiting}, and
RowHammer~\mpk{\cite{kim2014flipping, mutlu2017rowhammer, mutlu2019rowhammer,
walker2021on, yang2019trap, park2016statistical}}).
Section~\ref{position:subsubsec:design_characteristics} provides further detail
about such design properties and how knowing them is necessary to develop
effective test methodologies.}

\subsubsection{\mpl{Determination from Modeling}}

\mpl{Second, the system designer may make predictions from analytical or
empirical error models \circled{4} based on a previous understanding of DRAM
errors (e.g., from past experiments or scientific studies).} Examples of such
error models include: analytical models based on understanding DRAM failure
modes (e.g., sources of runtime faults~\mpk{\cite{croswell2000model, das2018vrl,
cardarilli2000development, hwang2012cosmic, siddiqua2013analysis,
meza2015large})}, parametric statistical models that provide useful summary
statistics (e.g., lognormal distribution of cell data-retention
times~\cite{hamamoto1995well, hamamoto1998retention, jin2005prediction,
hiraiwa1996statistical, li2011dram, hiraiwa1998local, edri2016silicon,
kim2009new, kong2008analysis}, exponential distribution of the time-in-state of
cells susceptible to variable-retention time
(VRT)~\mpm{\cite{bacchini2014characterization, kim2015avert, qureshi2015avatar,
yaney1987meta, restle1992dram, shirley2014copula, kim2011characterization,
kim2011study, kumar2014detection, mori2005origin, ohyu2006quantitative,
khan2014efficacy, kang2014co, liu2013experimental}}), physics-based simulation models (e.g.,
TCAD~\cite{yang2019trap, synopsys2018sentaurus, duan20172d, pfaffli2018tcad,
jin2005prediction} and SPICE models~\cite{luo2020clr, hassan2019crow,
lee2017design, hassan2016chargecache, lee2013tiered, shin2019dram,
wang2018reducing, zhang2016restore, wang2020figaro}), and empirically-determined
curves that predict observations well (e.g., single-bit error
rates~\mpk{\cite{patel2017reach, qureshi2015avatar, liu2013experimental,
khan2014efficacy, khan2016parbor, park2016statistical}}). \mpl{Similar to
testing, using error models to predict error characteristics ultimately relies
on understanding the DRAM chip being tested because the accuracy of the
predictions requires choosing appropriate models and model parameters (e.g.,
through testing \circled{3} or directly from fundamental chip design properties
\circled{1}).}

\subsection{\mpl{Access to Modeling and Testing Information}}
\label{position:subsec:knowing_or_assuming}

\mpl{Figure~\ref{fig:test_flow} shows that determining a DRAM chip's error
characteristics through modeling or testing ultimately relies on understanding
the chip's fundamental design properties. This reliance can be implicit (e.g.,
inherent within a pre-existing workflow designed for a specific chip) or
explicit (e.g., chosen as part of a home-grown testing methodology). Therefore,
a system designer must be vigilant of the information they (perhaps unknowingly)
rely upon at each step of their design process concerning commodity DRAM.}

\mpl{Fortunately, the system designer \emph{only} needs to be concerned with the
information flow at the children of a node whose information is already known
from a trustworthy source. For example, a system designer who wants to identify
the locations of error-prone cells (i.e., \circled{5}) using testing need not be
concerned with chip design properties (i.e., \circled{1}) if DRAM manufacturers
provide appropriate test methodologies (i.e., \circled{2}) or detailed test
results (i.e., \circled{3}).} Unfortunately, to our knowledge, neither DRAM
standards nor manufacturers provide the information in \emph{any} of the nodes
today, much less in a clear, industry-validated manner. \mpl{Therefore, the
system designer lacks a base of trustworthy information to build upon. This
creates a barrier to entry for a system designer who \mpm{wants} to explore
optimizations to commodity DRAM by compromising the designer's ability to make
well-informed \mpm{or effective} decisions.}

In general, except for the few major DRAM customers who may be able to secure
confidentiality agreements,\footnote{Even under confidentiality, DRAM
manufacturers may be unwilling to reveal certain proprietary aspects of their
designs (e.g., on-die error correction~\mpk{\cite{patel2020bit,
gurumurthi2021hbm3}}, target row refresh~\cite{saroiu2022price}) or provide
specifically requested numbers.} \mpl{system designers would need to rely on
(possibly incorrect \mpm{or incomplete}) \emph{inferences} or \emph{assumptions}
based on domain knowledge or reverse-engineering studies (e.g., similar in
spirit to~\mpm{\cite{patel2017reach, kraft2018improving, liu2013experimental,
patel2020bit, kim2020revisiting, jung2016reverse, barenghi2018software,
hassan2021uncovering, farmani2021rhat, frigo2020trrespass, wang2020dramdig,
pessl2016drama, jiang2021trrscope, kim2018solar, chang2016understanding,
lee2017design}}) that are not verified or supported by the DRAM
industry.}\footnote{DRAM manufacturers may make assumptions during their own
testing. However, they have full transparency into their own designs (i.e, the
root node in the \mpm{information} flow), so they can make the most informed
decision.} \mpk{As a result, the \emph{need} for assumptions can discourage
practitioners from exploring the full design space even when a given design
choice is otherwise beneficial. \mpl{We conclude that} the \emph{lack of
information transparency} is a serious impediment to adopting \mpm{many
promising} DRAM-related optimizations today.} 
\section{Study 1: Improving Memory Reliability}
\label{sec:rela_study}

\mph{Main memory reliability is a key design concern for \emph{any} system
because \emph{when} and \emph{how} memory errors occur affects overall system
reliability. In particular, designers of reliability-critical systems such as
enterprise-class computing clusters (e.g., cloud, HPC) and systems operating in
extreme or hostile environments (e.g., military, automotive, industrial,
extraterrestrial) take additional measures \mpk{(e.g., custom
components~\mpm{\cite{agrawal1994proposed, smart2021rugged, im2016im,
infineon2022radiation, lu1989advanced, banerjee1989two, mazumder1993design,
data2022rad, 3d2022ddr4}}, redundant resources~\cite{mathew2021thermoelectric,
kobayashi2017highly, patil2021dve})} to ensure that memory errors do not
compromise their systems. Section~\ref{subsubsec:mot_rela} shows the benefits of
incorporating mechanisms to improve memory reliability.} This section explains
how the details of a DRAM chip's reliability characteristics play a major role
in determining how system designers improve overall system reliability.

\subsection{Adapting Commodity DRAM Chips}
\label{subsec:rela_study_adapting_commodity_chips}

Commodity DRAM is designed to work for a wide variety of systems at a reasonable
(albeit unspecified)\footnote{\mph{Academic works speculate that commodity DRAM
targets a bit error rate (BER) within the range of
$10^{-16}-10^{-12}$~\cite{nair2013archshield, kim2016all,
longofono2021predicting, patel2017reach}, but we are unaware of
industry-provided values.}} error rate. \mph{In general, a system designer who
needs high memory reliability must design and build their own solutions (i.e.,
outside of the DRAM chip) to tolerate memory errors.\footnote{\mph{Even
designers who adopt custom DRAM solutions that sacrifice the cost advantages of
commodity memory (e.g., high-reliability DRAM~\cite{smart2021rugged, im2016im})
may supplement the DRAM chips with additional error-mitigation mechanisms
outside of the DRAM chip}.} In doing so, the designer effectively adapts a DRAM
chip to specific system needs, enhancing DRAM reliability beyond what the DRAM
chips provide alone.}

Section~\ref{subsubsec:mot_rela} reviews examples of such memory
error-mitigation mechanisms, which span the hardware-software stack. Regardless
of where each mechanism operates from, the mechanism targets a particular
\emph{error model}, which defines the scope of the errors that it is designed to
mitigate. This is important because, while a given mechanism efficiently
mitigates errors within its target error model, it may fail to do so if errors
no longer fit the model. In such cases, a different error-mitigation mechanism
\mpk{(or possibly, a combination of multiple mechanisms)} may be more suitable.

For example, a coarse-grained approach such as page
retirement~\mpk{\cite{mcelog2021bad, nvidia2020dynamic, venkatesan2006retention,
baek2014refresh, hwang2012cosmic, meza2015revisiting}} efficiently mitigates a
small number of errors at fixed bit positions. However, page retirement exhibits
significant capacity and performance overheads at high error rates or when
mitigating errors that change positions over time~\cite{meza2015revisiting,
lee2019exploiting, mcelog2021bad}. In contrast, a fine-grained hardware-based
approach such as a block error-correcting code~\mpk{\cite{moon2005error,
richardson2008modern, roth2006introduction, clark2013error, costello1982error,
costello2004ecc}} can efficiently mitigate a limited number of
randomly-distributed errors but can fail silently (and even exacerbate the
number of errors present~\mpm{\cite{alam2021lightweight, jeong2020pair,
criss2020improving, son2015cidra, patel2019understanding, patel2020bit,
patel2021enabling}}) when its correction capability is exceeded. We conclude
that it is essential for the system designer to \mpl{know} when and how errors
occur in a given memory chip in order to make an informed choice of which
error-mitigation mechanism to use in a particular system.

\subsection{\mpm{Lack of Transparency in Commodity DRAM}}
\label{subsec:rela_study_necessary_assumptions}

Unfortunately, system designers generally do not have access to definitive error
models for commodity DRAM chips. Therefore, designers are left to rely upon
information they can gather \mpk{by themselves} (e.g., by expending testing
resources) or from external, possibly untrustworthy, sources. However, as
Section~\ref{position:sec:formalizing} discusses, obtaining the error
characteristics of a DRAM chip without input from the manufacturers requires
making a series of assumptions about the chip's design and testing
methodologies. The need for these assumptions (i.e., the lack of trustworthy
information) can easily discourage designers from pursuing custom solutions to
enhance DRAM reliability. 

To exacerbate the problem of identifying a definitive error model, DRAM
manufacturers are starting to incorporate two on-die error-mitigation mechanisms
that correct a limited number of errors from within the DRAM chip itself: (1)
\emph{on-die ECC}~\cite{micron2017whitepaper, nair2016xed, oh2014a, kwak2017a,
kwon2014understanding, oh20153, patel2019understanding, patel2020bit} for
improving reliability and yield and (2) \emph{target row
refresh}~\cite{hassan2021uncovering, frigo2020trrespass, jattke2022blacksmith,
marazzi2022protrr} for partially mitigating the RowHammer vulnerability. Prior
works on ECC~\mpm{\cite{son2015cidra, gong2018duo, nair2016xed, jeong2020pair,
cha2017defect, pae2021minimal, criss2020improving, luo2014characterizing,
gurumurthi2021hbm3, patel2019understanding, patel2020bit, patel2021harp,
patel2021enabling}} and RowHammer~\mpm{\cite{saroiu2022price,
qureshi2021rethinking, frigo2020trrespass, hassan2021uncovering}} show that both
on-die ECC and TRR change how errors appear outside of the DRAM chip, thereby
changing the DRAM error model seen by the memory controller (and therefore, to
the rest of the system). \mpk{Unfortunately, both mechanisms are opaque to the
memory controller and are considered trade secrets that DRAM manufacturers will
not officially disclose~\cite{gong2017dram, nair2013archshield,
patel2019understanding, patel2020bit, saroiu2022price, qureshi2021rethinking,
loughlin2021stop, farmani2021rhat}. As a result, both on-die ECC and TRR make it
difficult for a system designer to reason about the DRAM error model and error
rates. For example, to account for on-die ECC's and TRR's effects when designing
a system-level error-mitigation mechanism, the system designer must spend
additional time and resources using reverse-engineering techniques (e.g., for
on-die ECC~\cite{patel2019understanding, patel2020bit} or
TRR~\mpm{\cite{hassan2021uncovering, frigo2020trrespass}}) or otherwise
\mpm{find} a trustworthy source to acquire the \mpm{necessary information in
reliable manner.}} 
\section{Study 2: DRAM Refresh Overheads}
\label{sec:refresh_study}

DRAM refresh is a key design concern \mpf{in modern systems.
Section~\ref{subsubsec:mot_dram_refresh} reviews evidence that reducing the
total number of refresh operations significantly benefits overall system
performance and energy efficiency}. In this section, we examine how mitigating
refresh overheads in commodity DRAM requires making assumptions about DRAM
reliability characteristics. Based on our analysis, we argue that these
assumptions limit the techniques' potential for adoption, discouraging system
designers from using these solutions in practice.

\subsection{Adapting Commodity DRAM Chips}
\label{subsec:adapting_commodity_dram_ref}

Reducing unnecessary refresh operations in commodity DRAM chips generally
requires two key steps. First, the memory controller must reduce the frequency
of periodic refresh operations. This is \mpk{achievable (though not necessarily
supported to arbitrary values) using commodity DRAM chips because the memory
controller manages DRAM refresh timings. For example, the memory controller
might} relax the rate at which it issues refresh operations \mpk{to half of the
DDR\emph{n} standard of 3.9 or 7.8 $\mathrm{\mu}$s, which is supported by
standards at extended temperature ranges~\cite{jedec2020ddr5,
jedec2020lpddr5,jedec2014lpddr4, jedec2012ddr4, jedec2008ddr3}, or even to over}
an order of magnitude less often~\cite{venkatesan2006retention, liu2012raidr,
nair2013archshield, katayama1999fault}.

Second, the system must mitigate any errors that may occur within the small
number of DRAM cells that require frequent refreshing. Doing so requires either
using additional refresh operations (e.g., by issuing extra row
activations~\cite{liu2012raidr}) or using error-mitigation mechanisms within
processor (e.g., ECC~\cite{qureshi2015avatar} and/or bit-repair
techniques~\cite{venkatesan2006retention, lin2012secret, nair2013archshield}).
Although both strategies introduce new performance and energy overheads,
\mpl{the benefits of reducing unnecessary refresh operations outweigh the
overheads introduced~\cite{liu2012raidr, ohsawa1998optimizing,
wang2014proactivedram, venkatesan2006retention, lin2012secret,
nair2013archshield, ghosh2007smart, qureshi2015avatar, patel2017reach,
nguyen2021zem}. For example, Liu et al.~\cite{liu2012raidr} project that DRAM
refresh overheads cause a 187.6\% increase in the energy-per access and a 63.7\%
system performance degradation for 64~Gib chips. By reducing the overall number
of DRAM refresh operations, the authors show that their mechanism, RAIDR, can
mitigate these overheads by 49.7\% and 107.9\%, respectively.} 

\subsection{\mpm{Lack of Transparency in Commodity DRAM}}
\label{subsec:limitng_assumptions_dram_ref}

Knowing, predicting, or identifying cells that cannot safely withstand
infrequent refreshing (i.e., retention-weak cells) is a difficult reliability
\mpk{problem because the cells' likelihood of error changes with how a DRAM chip
is used (i.e., operating conditions such as the refresh rate, voltage,
temperature) and the particular DRAM chip circuit design (e.g., random
cell-to-cell variations, locations of true and
anti-cells~\cite{liu2013experimental, kraft2018improving,
patel2019understanding}). Prior works propose two practical ways of identifying
retention-weak cells: (1) \emph{active profiling}, which uses comprehensive
tests to search for error-prone cells offline~\cite{liu2012raidr,
patel2017reach, khan2016parbor, khan2014efficacy, lin2012secret,
mathew2017using}, and (2) \emph{reactive profiling}, which constantly monitors
memory to identify errors as they manifest during runtime, e.g., ECC
scrubbing~\cite{qureshi2015avatar, han2014data, choi2020reducing}. Both
approaches require the profiler to understand the \emph{worst-case} behavior of
data-retention errors for a given DRAM chip~\cite{khan2014efficacy,
lin2012secret}: an active profiler must use the worst-case conditions to
maximize the proportion of retention-weak cells it identifies during
profiling~\cite{patel2017reach} and a reactive profiler must be provisioned to
identify (and possibly also mitigate) the worst-case error pattern(s) that might
be observed at runtime, e.g., to choose an appropriate ECC detection and
correction capability~\cite{patel2021harp, qureshi2021rethinking,
khan2014efficacy}.}

\mpk{The fact that an effective error profiling mechanism relies on
understanding the underlying error characteristics reinforces the argument
presented in Section~\ref{position:sec:formalizing}. Even though there exist
techniques for mitigating refresh overheads in commodity DRAM, practically
adopting them relies on prerequisite knowledge about a DRAM chip and its
reliability characteristics that is not provided by the DRAM industry
today.} 
\section{Study 3: Long DRAM Access Latency}
\label{sec:latency_study}

Slow generational improvements in the DRAM access latency \mpf{(shown in
Section~\ref{sec:intro}) contrast with the growing prevalence of
latency-sensitive workloads today~\mpm{\cite{hsieh2016accelerating,
ferdman2012clearing, gutierrez2011full, hestness2014comparative, huang2014moby,
zhu2015microarchitectural, oliveira2021damov, boroumand2018google,
boroumand2021google, koppula2019eden, kanellopoulos2019smash, son2013reducing,
mutlu2013memory, wilkes2001memory, wulf1995hitting, mutlu2007stall,
mutlu2003runahead, kanev2015profiling, mutlu2014research, bera2019dspatch,
bera2021pythia, liu2019binary, ghose2019processing, shin2014nuat,
ghose2019demystifying, gomez2021benchmarking, gomez2021benchmarkingmemory,
giannoula2022towards}}. Therefore, as Section~\ref{subsubsec:mot_dram_latency}
discusses, there is significant opportunity for improving overall system
performance by reducing the memory access
latency~\mpk{\cite{chandrasekar2014exploiting, chang2016understanding,
kim2018solar, lee2015adaptive, lee2017design, wang2018reducing,
zhang2016restore, hassan2016chargecache, koppula2019eden, mathew2017using,
zhang2021quantifying, kim2020improving, lee2016simultaneous}}.} In this section,
we study how \mpk{techniques} for reducing the access latency of commodity DRAM
chips rely on making assumptions about DRAM reliability characteristics. Then,
we argue that \mpk{the need for these assumptions \mpm{(and the lack of
transparency in DRAM to allow them)} discourages system designers from adopting
the latency reduction techniques}.

\subsection{Adapting Commodity DRAM Chips}
\label{subsec:adapting_commodity_dram_lat}

Strategies for improving the access latency of commodity DRAM chips rely on
manipulating DRAM commands and/or access timings to either (1) \mpk{eliminate}
conservative timing margins that DRAM manufacturers use to account for
worst-case operation~\mpm{\cite{chang2016understanding, lee2015adaptive,
chandrasekar2014exploiting, mathew2017using, shin2014nuat, kim2018solar,
hassan2016chargecache, david2011memory, deng2011memscale,
zhang2021quantifying}}; or (2) exploit undefined DRAM chip behavior to perform
beneficial operations (e.g., performing massively-parallel computations within
DRAM rows~\mpm{\cite{hajinazar2021simdram, gao2019computedram,
seshadri2013rowclone, olgun2021pidram, seshadri2017ambit, seshadri2017simple,
seshadri2015fast, seshadri2016buddy, seshadri2020indram}}, generating random
values~\cite{kim2019d, talukder2018exploiting, olgun2021quac} or unique chip
identifiers~\cite{kim2018dram, talukder2018ldpuf, yue2020dram,
hashemian2015robust, schaller2018decay}). 

In both cases, \emph{new} DRAM access timings must be \mpk{determined that}
ensure the desired operation can be performed predictably and reliably
\mpk{under all conditions}. To identify \mpk{these access} timings, prior
works~\mpk{\cite{kim2018dram, kim2019d, talukder2019prelatpuf,
gao2019computedram, olgun2021quac, chang2016understanding,
chang2017understanding, lee2015adaptive, chandrasekar2014exploiting,
ghose2018your, hassan2017softmc, chang2017thesis, lee2016reducing,
david2011memory} perform extensive experimental characterization studies across
many DRAM chips.} These studies account for three primary sources of variation
that affect the access timings of a given memory location. First, process
variation introduces random variations between DRAM chip components (e.g.,
cells, rows, columns). Second, a manufacturer's particular circuit design
introduces structural variation (called design-induced
variation~\cite{lee2017design}) that deterministically affects access timings
based on a component's location in the overall DRAM design (e.g., cells along
the same bitline~\cite{kim2018solar}, cells at the borders of internal storage
arrays~\cite{lee2017design}). Third, the charge level of a DRAM cell varies over
time due to leakage and the effects of DRAM accesses~\cite{shin2014nuat,
hassan2016chargecache}. Experimentally determining the \mpk{new predictable and
reliable} access timings requires \mpk{properly} accounting for all three
sources of variation \mpk{under all operating conditions}.

\subsection{\mpm{Lack of Transparency in Commodity DRAM}}
\label{subsec:limitng_assumptions_dram_lat}

Unfortunately, determining new viable access timings requires developing and
executing a reliable testing methodology, which in turn requires making similar
assumptions to those discussed for \mpk{data-retention error profiling in
Section~\ref{subsec:limitng_assumptions_dram_ref}. Choosing runtime (e.g., data
and access patterns) and environmental (e.g., temperature, voltage) testing
conditions \emph{in a meaningful way} requires some understanding of the error
mechanisms involved in timing-related errors~\cite{khan2017detecting}, including
(but not limited to)} aspects of the circuit design, such as internal
substructure dimensions (e.g., subarray sizing)~\cite{kim2018solar,
lee2017design}, the correspondence between logical DRAM bus addresses and
physical cell locations~\mpm{\cite{chang2016understanding, lee2015adaptive,
khan2016parbor}}, and the order of rows refreshed by each auto-refresh
operation~\cite{shin2014nuat}. \mpk{A system designer is discouraged from
exploring improvements to the commodity DRAM access latency without trustworthy
access to this information.} 
\section{Study 4: RowHammer Mitigation}
\label{sec:security_study}

\mpf{Many promising proposals exist for adding RowHammer defenses to commodity
DRAM chips} (discussed in Section~\ref{subsubsec:mot_dram_security}), but their
potential for adoption is hampered by system designers' lack of visibility
\mpf{into how the underlying error mechanism behaves.} In this section, we
examine the various assumptions that RowHammer defense proposals rely upon and
argue that these assumptions pose serious barriers for practical adoption.

\subsection{Adapting Commodity DRAM Chips}

\mpf{To effectively mitigate RowHammer bit flips, a mitigation mechanism must be
configured based on the vulnerability level of a given DRAM chip. This requires
estimating the chip's RowHammer error characteristics for different operating
conditions and access patterns. Each of the four mechanism types introduced in
Section~\ref{subsubsec:mot_dram_security} requires estimating different
characteristics.} Table~\ref{tab:rh_mitigation_info} summarizes the different
pieces of information required for each mitigation type. The first is known as
\hcfirst{}~\mpk{\cite{kim2020revisiting, orosa2021deeper} or RowHammer
Threshold~\cite{kim2014flipping, yaglikci2020blockhammer,
bennett2021panopticon}}, which describes the worst-case number of RowHammer
memory accesses required to induce a bit-flip. The second is known as the blast
radius~\cite{kim2014flipping, kim2020revisiting}, which describes how many rows
\mpf{are affected by hammering a single row}. The third is the DRAM's internal
physical row address mapping~\mpm{\cite{kim2014flipping, kim2012case}}, which is
necessary to identify the locations of victim rows.

\begin{table}[h]
    \centering
    \small
    \begin{tabular}{l|ccc}
                           & \multicolumn{3}{c}{\textbf{Required Information}} \\
        \textbf{Strategy}  & \hcfirstbold{}  &  \textbf{Blast Radius} &  \textbf{Row Mapping } \\\hline 
        Access-Agnostic       & \cmark & & \\               
        Proactive             & \cmark & \cmark & \\         
        Physically Isolating  & \cmark & \cmark & \cmark \\                  
        Reactive              & \cmark & \cmark & \cmark
    \end{tabular}
    \caption{Information needed by each of the four RowHammer-mitigation strategies.}
    \label{tab:rh_mitigation_info}
\end{table}

All three RowHammer error characteristics vary between DRAM manufacturers,
chips, \mpk{and cells} based on a combination of random process variation, a
manufacturers' particular circuit design (including yield-management techniques
such as post-manufacturing repair, target row refresh, and error correcting
codes), \mpk{and operating conditions such as temperature and
voltage~\cite{kim2014flipping, park2016statistical, park2016experiments,
kim2020revisiting, orosa2021deeper, yaglikci2021security, yun2018study,
lim2016active, farmani2021rhat}. Therefore, as with estimating DRAM refresh and
access timings (discussed in Sections~\ref{subsec:limitng_assumptions_dram_ref}
and~\ref{subsec:limitng_assumptions_dram_lat}), these studies rely on extensive
experimental testing to estimate RowHammer error characteristics that are needed
to design and/or configure the RowHammer defenses discussed in
Section~\ref{subsubsec:mot_dram_security}.}

\subsection{\mpm{Lack of Transparency in Commodity DRAM}}
\label{position:subsec:rh:necessary}

We observe that \mpk{all} previously-proposed RowHammer mitigation mechanisms
require accurately estimating RowHammer error characteristics \mpk{throughout
all valid operating conditions}. In particular, \mpm{every mechanism} must be
tuned against \mpg{at least} \hcfirst{} in order to effectively prevent
RowHammer. \mpm{Prior works~\mpk{\cite{qureshi2021rethinking, saroiu2022price,
loughlin2021stop}} make the same observation, discussing the difficulty in
practically determining and relying on this information without support from
DRAM manufacturers.} 

\mpm{Therefore, a security-focused system designer who wants to implement or
build upon one of the many previously-proposed system-level RowHammer defense
mechanisms (discussed in Section~\ref{subsubsec:mot_dram_security}) is limited
by the same information access challenges as discussed in
Section~\ref{position:subsec:knowing_or_assuming}: because neither the error
characteristics they need nor the methods to obtain them are provided by
official sources, the system designer must rely on other means to obtain the
necessary information. As a result, the system designer is likely discouraged
from exploring designs that} \mpn{address RowHammer errors in commodity DRAM
chips altogether.} 
\section{\mpk{Current} DRAM Standards as the Problem}
\label{sec:spec_as_problem}

Based on our case studies, we conclude that reliance on information about DRAM
reliability characteristics poses a serious challenge for optimizing how
commodity DRAM is used. In this section, we hypothesize that the unavailability
of \mpk{information related to DRAM reliability} is caused by a \emph{lack of
transparency} within DRAM standards which provide \emph{control over}, but not
\emph{insight into}, DRAM operations. We identify DRAM standards as both (1) the
root cause of having to make assumptions about DRAM reliability \mpk{(as
standards are currently defined)} and (2) the pathway to a solution for
alleviating the need for such assumptions \mpk{(by incorporating DRAM
reliability as a key concern)}.

\subsection{\xmt{The Problem of Information Unavailability}}
\label{subsec:problem_of_unavailability}

\mpk{In each case study} throughout
Sections~\ref{sec:rela_study}--\ref{sec:security_study}, we observe that
optimizing commodity DRAM chips for key system design concerns requires knowing
information about DRAM reliability. This is unsurprising because
\mpk{reliability is central to each case study's approach: each study improves
system-level metrics (e.g., reliability, energy-efficiency, performance,
security) by leveraging key properties of one or more error mechanisms (e.g.,
spatiotemporal dependence of errors due to circuit timing
violations~\cite{kim2018solar, chang2016understanding, lee2017design}, the
localized nature of RowHammer errors~\mpm{\cite{kim2014flipping,
konoth2018zebram, van2018guardion, brasser2017can, kim2020revisiting}}).}
Therefore, identifying the best operating point requires \mpk{at least a basic
understanding} of how the error mechanisms themselves behave under
\mpk{representative} operating conditions.

Recent works~\cite{qureshi2021rethinking, saroiu2022price} discuss the pitfalls
of \mpk{designing defense mechanisms that rely on knowledge of how RowHammer
errors behave (e.g., \hcfirst{}, dependence on a chip's internal cell
organization)}, calling into question the practicality of accurately determining
\mpk{these details given an arbitrary DRAM chip.} Knowing or determining this
information is essential to guarantee protection against RowHammer. However,
determining it \mpk{without guidance from DRAM manufacturers} requires per-chip
testing and/or reverse-engineering that relies on the accuracy of the underlying
testing methodology used, which itself relies on knowledge of \mpk{DRAM chip
details} that likely needs to be assumed or inferred (as discussed in
Sections~\ref{position:sec:formalizing} and~\ref{position:subsec:rh:necessary}). 

As a result, a system designer who wants to adapt commodity DRAM for their
design requirements today is forced to make design and/or mechanism
configuration decisions based upon assumptions or inferences from unofficial
sources (e.g., self-designed experimental studies~\mpm{\cite{lee2015adaptive,
lee2017design, kim2018solar, kim2018dram, kim2019d, talukder2018exploiting,
talukder2019prelatpuf, talukder2018ldpuf, chang2016understanding,
kim2014flipping, liu2013experimental, hassan2017softmc,gao2019computedram,
olgun2021quac, chang2017understanding, chandrasekar2014exploiting,
ghose2018your, chang2017thesis, lee2016reducing, david2011memory}}).
Unfortunately, even a system designer willing to spend significant resources on
such adaptations (e.g., \mpk{to enhance system} reliability, performance,
security, etc.) may be discouraged by the underlying dependence on untrustworthy
information. In the worst case, \mpk{the designer may judge} \emph{all}
adaptations \mpk{to be} impractical without a trustworthy understanding of a
DRAM chip. \mpk{We conclude that the lack of information transparency today
discourages system designers} from exploring alternative designs that have been
shown to provide tangible benefits.

\begin{table*}[b]
    \centering
    \small
    \setlength\tabcolsep{3pt}
    \begin{tabular}{L{5cm}|L{3.5cm}L{9cm}}
        \textbf{Design Characteristic} & \textbf{Reverse-Engineered By} & \textbf{Use-Case(s) Relying on Knowing the Characteristic} \\\hline\hline
        \begin{tabular}[c]{@{}l@{}}Cell charge encoding convention\\(i.e., true- and anti-cell layout)\end{tabular} & Testing~\cite{patel2019understanding, kraft2018improving, patel2017reach, liu2013experimental} & Data-retention error modeling and testing \mpk{for mitigating refresh overheads} (e.g., designing worst-case test patterns)~\cite{khan2017detecting, kraft2018improving, liu2013experimental} \\\hline
        On-die ECC details & Modeling and testing~\cite{patel2019understanding, patel2020bit} & Improving reliability (e.g., designing ECC within the memory controller)~\cite{son2015cidra, cha2017defect, criss2020improving, gong2018duo}, \mpk{mitigating RowHammer}~\cite{kim2020revisiting, hassan2021uncovering, cojocar2019exploiting, jattke2022blacksmith}  \\\hline
        Target row refresh (TRR) details & Testing~\cite{hassan2021uncovering, frigo2020trrespass} & Modeling and mitigating RowHammer~\cite{frigo2020trrespass, hassan2021uncovering, jattke2022blacksmith} \\\hline
        \mpk{Mapping between internal and external row addresses} & Testing~\cite{jung2016reverse, kim2020revisiting, tatar2018defeating, barenghi2018software, wang2020dramdig, lee2017design} & Mitigating RowHammer~\cite{kim2014flipping, barenghi2018software, jung2016reverse, kim2020revisiting, farmani2021rhat} \\\hline
        \mpk{Row addresses refreshed by each refresh operation} & Testing~\cite{hassan2021uncovering} & Mitigating RowHammer~\cite{hassan2021uncovering}, improving access timings~\cite{shin2014nuat, wang2018reducing} \\\hline
        Substructure organization (e.g., cell array dimensions) & Modeling~\cite{lee2017design} and testing~\cite{chang2016understanding, lee2017design, kim2018solar} & Improving DRAM access timings~\cite{lee2017design, chang2016understanding, kim2018solar} \\\hline
        \begin{tabular}[c]{@{}l@{}}Analytical model parameters\\(e.g., bitline capacitance)\end{tabular} & Modeling and testing~\cite{hamamoto1998retention, liu2013experimental} & \mpm{Developing and using error models for improving overall reliability~\cite{li2011dram}, mitigating refresh overheads (e.g.,~data-retention~\cite{das2018vrl, hamamoto1998retention, hiraiwa1996statistical} and VRT~\cite{restle1992dram, shirley2014copula} models), improving access timings~\cite{lee2017design}, and mitigating RowHammer~\cite{walker2021dram, park2016statistical}} \\
    \end{tabular}
    \caption{Basic DRAM chip design characteristics that are typically assumed or inferred for experimental studies.}
    \label{tab:design_characteristics}
\end{table*}

\subsection{Limitations of DRAM Standards}

Current DRAM standards do not address general reliability characteristics
because commodity DRAM is designed for a fixed, high-reliability operating point
such that the typical consumer can largely ignore errors. This follows directly
from the separation-of-concerns between system and DRAM designers: current DRAM
standards place \mpg{most of the burden of addressing DRAM} reliability
challenges (e.g., worsening error rates with continued technology
scaling~\mpk{\cite{kang2014co, micron2017whitepaper, mutlu2013memory}) on DRAM
manufacturers} alone.\footnote{\mpg{High-reliability systems may supplement DRAM
chips' base reliability with additional error-mitigation mechanisms, as
discussed in Section~\ref{subsubsec:mot_rela}.}}

We believe that this state of affairs arises naturally because establishing a
strict separation of concerns requires a clear and explicit interface between
manufacturers and customers. Consequently, ensuring that the standards leave
enough flexibility for diverse customer use-cases requires careful and explicit
attention. This is because the standards are susceptible to \emph{abstraction
inversion}~\cite{baker1990opening}, a design anti-pattern in which a previously
agreed-upon interface becomes an \emph{obstacle}, forcing \mpk{system designers}
to re-implement basic functionality in terms of the outdated abstraction. A
rigid interface limits what is and is not possible, potentially requiring
unproductive reverse-engineering to work around.

We argue that \mpk{needing to make assumptions in order to adapt} commodity DRAM
to system-specific goals \mpk{clearly indicates} abstraction inversion today.
This implies that DRAM standards have aged without sufficient attention to
flexibility. Although a fixed operating point defines a clear interface, we
believe that leaving room for (and potentially even encouraging) different
operating points \mpk{is essential today.}
 
\section{DRAM \mpg{Standards} as the Solution}
\label{sec:two_part_change_to_specs}

We believe that the separation of concerns provided by DRAM standards is
necessary for practicality because it enables DRAM manufacturers and system
designers to focus on designing the best possible products within their
respective areas of expertise. However, we argue that the separation must be
crafted in a way that not only does not impede progress, but ideally encourages
\mpk{and aids} it. To achieve both goals, we propose extending DRAM standards in
a way that enables system designers to make informed decisions about how their
design choices will affect DRAM operation. In other words, instead of modifying
DRAM \emph{designs}, we advocate modifying \emph{standards} to facilitate
transparency of DRAM reliability characteristics. Armed with this information,
system designers can freely explore how to best use commodity DRAM chips to
solve their own design challenges while preserving the separation of concerns
that allows DRAM designers to focus on building the best possible
standards-compliant DRAM chips.

\subsection{\mpl{Choosing Information to Release}}
\label{position:subsec:what_to_release}

\mpg{We identify what information to release using our analysis of information
flow in Section~\ref{position:sec:formalizing}. We observe that, given the
information at \emph{any} node in Figure~\ref{fig:test_flow}, system designers
can work to determine the information at each of its child nodes. As a result,
\mpi{access to trustworthy} information at \emph{any} node \mpi{provides system
designers with a foundation to make informed design decisions.} Therefore, we
recommend that the DRAM industry be free to release information at \emph{at
least one} node of their choice that they are willing and capable of doing so.
This} section examines realistic possibilities for communicating information at
each \mpg{node} of the flowchart.

\subsubsection{\xmt{Basic Design Characteristics}}
\label{position:subsubsec:design_characteristics}

At the lowest level, DRAM manufacturers could provide basic chip design
characteristics that allow system designers to develop their own test
methodologies and error models. This is the most general and flexible approach
because it places no limitations on what types of studies system designers may
pursue (e.g., in contrast to providing information \mpk{that is useful for
reasoning about only one} particular error mechanism).
Table~\ref{tab:design_characteristics} gives examples of key design
characteristics that prior works often make assumptions about in their own
efforts to optimize commodity DRAM usage. \mpk{For each design characteristic,
we list prior works that reverse-engineer the characteristic and describe
use-cases that rely on knowledge of the characteristics.}

We believe that releasing \mpg{these characteristics will minimally (if at all)}
impact DRAM manufacturer's business interests given that each of the
characteristics can be reverse-engineered with existing methods \mpk{(as shown
by Table~\ref{tab:design_characteristics}, Column 2)} and access to appropriate
tools, as demonstrated by prior studies~\mpm{\cite{patel2019understanding,
kraft2018improving, patel2017reach, liu2013experimental, patel2020bit,
hassan2021uncovering, frigo2020trrespass, jung2016reverse, kim2020revisiting,
lee2017design, chang2016understanding, kim2018solar, hamamoto1998retention,
barenghi2018software, wang2020dramdig, mukhanov2020dstress, kim2014flipping,
khan2014efficacy, farmani2021rhat}}. Releasing this information in an official capacity simply
confirms what is already suspected, providing a competitor with no more
information about a given DRAM chip than they already had available. On the
other hand, knowing this information empowers system designers \mpk{and enables
them to confidently design and implement system-level optimizations}, benefiting
both designers and manufacturers in the long run (as discussed in
Section~\ref{sec:motivation_new}).

\subsubsection{Test Methodologies}

\mpk{At a level of abstraction beyond chip design details}, DRAM manufacturers
could describe effective test methodologies that system designers can use to
study the particular aspects of DRAM reliability they are interested in.
Compared with providing chip design characteristics, directly providing test
methodologies absolves (1) manufacturers from needing to reveal chip design
information; and (2) system designers from needing the DRAM-related expertise to
determine the test methodologies from chip design characteristics.\footnote{We
believe that interested parties already have such expertise, as shown by the
fact that many studies~\mpm{\cite{patel2019understanding, kraft2018improving,
patel2017reach, liu2013experimental, patel2020bit, hassan2021uncovering,
frigo2020trrespass, jung2016reverse, kim2020revisiting, lee2017design,
chang2016understanding, kim2018solar, hamamoto1998retention,
barenghi2018software, wang2020dramdig, mukhanov2020dstress, kim2014flipping,
khan2014efficacy, farmani2021rhat}} determine the necessary test methodologies through extensive
experimentation.} \mpk{As a drawback}, providing test methodologies alone limits
system designers to working with only the particular error mechanisms that the
methodologies are designed for (e.g., data-retention, RowHammer).
Table~\ref{tab:test_params} summarizes key aspects of testing methodologies that
prior works generally need to assume throughout the course of their testing.

\begin{table}[t]
    \centering
    \small
    \begin{tabular}{L{2.2cm}|L{5.8cm}}
        \textbf{Test Parameter} & \textbf{Description} \\\hline\hline
        Data pattern 
            & Data pattern that maximizes the chance of errors occurring~\mpk{\cite{duganapalli2016modelling, liu2013experimental, khan2014efficacy, kim2014flipping, mukhanov2020dstress, patel2017reach, kim2020revisiting, orosa2021deeper,hassan2021uncovering, cojocar2020are,tatar2018defeating, cojocar2021mfit, jattke2022blacksmith, borucki2008comparison, weis2015retention, kraft2018improving, mukhanov2020dstress,kim2018solar}} \\\hline
        Environmental conditions 
            & Temperature and voltage that lead to worst-case behavior~\mpk{\cite{park2016experiments, kim2019d, liu2013experimental, yaglikci2022understanding, orosa2021deeper, schroeder2009dram, weis2015retention,wang2018dram, yaney1987meta,hamamoto1998retention}} \\\hline
        Test algorithm 
            & Sequence of representative and/or worst-case DRAM operations to test~\mpk{\cite{cojocar2020are, kim2014flipping, liu2013experimental, lee2015adaptive, kim2018dram, kim2019d, hassan2021uncovering, jattke2022blacksmith, salman2021half}}
    \end{tabular}
    \caption{Testing parameters that are typically assumed or inferred during experimental studies.}
    \label{tab:test_params}
\end{table}

\subsubsection{Test Results and/or Error Models}
\label{position:subsubsection:test_results}

\mpk{At the highest level of abstraction}, DRAM manufacturers can directly
provide test results and/or error models related to specific studies needed by
system designers. This could take the form of parametric error models (e.g., the
statistical relationship between operating timings and error rates) along with
parameter values for each chip, fine-granularity error characteristics (e.g.,
per-column minimum viable access timings) and/or summary statistics of interest
(e.g., \hcfirst{} in studies pertaining to RowHammer). In this way, system
designers can constrain (or entirely bypass) testing when developing mechanisms
using the provided information. \mpk{As a drawback}, directly releasing test
results and/or error models constrains system designers to developing solutions
only for those design concerns that pertain to the released information.
Table~\ref{tab:test_result_error_model} provides examples of key test results
and error models that prior works \mpk{leverage in order to implement
optimizations to commodity DRAM.}

\begin{table}[h]
    \centering
    \small
    \setlength\tabcolsep{3pt}
    \begin{tabular}{L{2.2cm}L{6cm}}
        \textbf{Test Result or Error Model} & \textbf{Description} \\\hline\hline
        Data-retention times
            & Minimum refresh rate required for different DRAM regions (e.g., rows, cells)~\mpk{\cite{liu2012raidr, lin2012secret, liu2013experimental,khan2014efficacy, khan2016case, kim2001block, nair2013archshield}} \\\hline
        Error profile
            & \mpk{List of cells susceptible to errors (e.g., VRT~\cite{qureshi2015avatar, liu2013experimental, khan2014efficacy}, latency-related~\cite{kim2019d, kim2018dram, kim2018solar, chang2016understanding, chandrasekar2014exploiting})} \\\hline
        Error rate summary statistics
            & \mpk{Aggregate error rates (e.g., BER~\cite{liu2013experimental, patel2017reach, patel2019understanding, weis2015retention,kang2014co}, FIT~\cite{schroeder2009dram, levy2018lessons, wang2009soft}), distribution parameters (e.g., copula~\cite{shirley2014copula}, lognormal~\cite{hamamoto1995well, hamamoto1998retention, li2011dram}, exponential~\cite{liu2012raidr,kumar2014detection})} \\\hline
        RowHammer blast radius
            & Maximum number of rows affected by hammering \mpm{one or more row(s)}~\mpk{\cite{yaglikci2020blockhammer, kim2020revisiting, yaglikci2021security, walker2021dram, loughlin2021stop, kim2014flipping}} \\\hline
        \hcfirst{} or RowHammer Threshold
            & \mpk{Minimum number of RowHammer accesses required to induce bit-flips~\cite{kim2020revisiting, orosa2021deeper, kim2014flipping, yaglikci2020blockhammer, bennett2021panopticon}}
    \end{tabular}
    \caption{\mpg{Examples of key test results and error models from prior works that study and/or optimize commodity DRAM.}}
    \label{tab:test_result_error_model}
\end{table}

\subsection{\mpl{Choosing} When to Release the Information}

\mpi{We expect that releasing information by changing DRAM standards will be a
slow process due to the need for consensus between DRAM stakeholders. Instead,
we propose decoupling the \emph{release} of information from the
\emph{requirement} to do so. To this end, we recommend a practical two-step
process with different approaches in the short- and long-term.}

\subsubsection{Step 1: \mpg{Immediate Disclosure of Information}}

\mpg{We \mpi{recommend} two independent approaches to quickly release
information in the short-term. First, we recommend a public crowdsourced database
\mpi{that aggregates already-known information, e.g., inferred through}
reverse-engineering studies. We believe this is practical given the significant
research and industry interest in optimizing how commodity DRAM chips are used.
Such a database would provide an opportunity for peer review of posted
information, increasing the likelihood that the information is trustworthy. In
the long run, we believe such a database would facilitate information release
from DRAM manufacturers themselves because the manufacturers could simply
validate database information, if not contribute directly.}

\mpg{Second, we recommend} that commodity DRAM manufacturers individually
release one or more of the aforementioned categories of information for current
DRAM chips and those already in the field. For example, manufacturers may
\mpi{update} chip datasheets to incorporate relevant design characteristics or
make more extensive information available online (e.g., similar to how some
manufacturers already provide compliance documents and functional simulation
\mpk{models} through their websites~\mpk{\cite{micron2021dram, issi2022ddr4,
nanya2022NT5AD256M16E4}}). Releasing \mpk{any of the information described
throughout Section~\ref{position:subsec:what_to_release}} requires no changes to
DRAM designs or standards, though modifying DRAM standards (e.g., via an
addendum, as we suggest in Step 2) would help unify the information release
across all manufacturers. However, in the short term, we believe it is more
important to release the information, even if \mpk{not} standardized, so that it
is available as soon as possible.

\subsubsection{Step 2: Explicit DRAM Reliability Standards}

In the long term, we recommend DRAM standards be modified to \mpk{promote (or
even require)} DRAM manufacturers to disclose any information that impacts DRAM
reliability as relevant to a system designer. This information may include any
or all of the information discussed throughout this work; we believe that the
DRAM stakeholders themselves (i.e., DRAM manufacturers and system designers) are
in \mpk{a good} position to determine and standardize which information is the
most relevant and useful to regulate.

As a concrete example of how such changes to standards may occur, we reference
test methodologies~\cite{jedec2010ssdrequirements, jedec2010ssdendurance} and
error models~\cite{jedec2016failure} that JEDEC provides for NAND \mpm{flash
memory endurance~\cite{cai2017error, cai2018errors, cai2012error}, including
floating-gate data retention~\cite{cai2015data, luo2018heatwatch,
luo2018improving, cai2012flash} and threshold voltage
distributions~\cite{cai2013threshold, cai2013program, cai2015read,
luo2016enabling}}. These documents outline standardized best practices for
studying and characterizing endurance properties of SSD devices. We envision
analogous documents released for key DRAM error mechanisms (e.g.,
data-retention, access-timing-related, RowHammer), providing a standardized and
reliable alternative to inferring the same information through unofficial
channels.

\subsection{Alternative Futures}

We anticipate consumer use-cases \mpk{to continue diversifying}, making
affordable-yet-flexible DRAM increasingly important. Ambitious initiatives such
as DRAM-system co-design~\mpk{\cite{patterson1997case, mutlu2014research,
mutlu2021primer, kim2014flipping, mutlu2015main}} and emerging, non-traditional
DRAM architectures~\mpm{\cite{devaux2019true, kwon202125, mutlu2021primer,
oliveira2021damov, he2020newton, niu2022184qps, ahn2016scalable, lee20221ynm,
mutlu2019processing, gomez2021benchmarking, gomez2021benchmarkingmemory}} will naturally engender transparency by tightening the
relationship between DRAM manufacturers and system designers. Regardless of the
underlying motivation, we believe that increased transparency of DRAM
reliability characteristics will remain crucial to allowing system designers to
make the best use of commodity DRAM chips \mpk{by enabling them to customize
DRAM chips for system-level goals}. 
\section{Conclusion}
\label{sec:conclusion}

We \mpm{contend} that system designers lack the necessary transparency into DRAM
reliability to make informed decisions about how their design choices will
affect DRAM operation. Without this transparency, system designers are
discouraged from exploring \mpk{the full design space around commodity DRAM,
wasting considerable potential for system-level optimization in meeting the
particular needs of their systems. We support our argument with} four case
studies that each examine an important design concern in modern DRAM-based
systems: (1) improving DRAM reliability; (2) mitigating DRAM refresh overheads;
(3) decreasing the DRAM access latency; and (4) defending against RowHammer. For
each case study, we argue that developing an effective system-level solution
requires making restrictive, potentially incorrect assumptions about DRAM
reliability characteristics. Based on our studies, we identify DRAM standards as
the source of the problem: current standards enforce a fixed operating point
without providing the context necessary to enable safe operation outside that
point. To overcome this problem, we introduce a two-step approach that modifies
DRAM standards to incorporate transparency of key reliability characteristics.
We believe that our work paves the way for a more open and flexible DRAM
standard that \mpk{enables \mpm{DRAM} consumers to better adapt and build upon
commodity DRAM technology while allowing \mpm{DRAM} manufacturers to preserve
their competitive edge. As a result, our work enables better innovation of
customized DRAM systems to fully harness the advantages of DRAM technology into
the future.} 

\section*{Acknowledgments}

\mpl{We thank the members of the SAFARI \mpm{Research Group} for their valuable
feedback and the constructively critical environment that they provide. We
specifically thank Geraldo F. Oliveira, Jisung Park, Haiyu Mao, Jawad Haj-Yahya,
Jeremie S. Kim, Hasan Hassan, Joel Lindegger, and Meryem Banu Cavlak for the
feedback they provided on earlier versions of this paper. We thank external
experts who helped shape our arguments, including Mattan Erez, Moinuddin
Qureshi, Vilas Sridharan, and Christian Weis. We acknowledge the generous gifts
provided by our industry partners, including Google, Huawei, Intel, Microsoft,
and VMware. We acknowledge support from the ETH Future Computing Laboratory and
the Semiconductor Research Corporation. This work is part of Minesh Patel's
Ph.D. Dissertation~\cite{patel2021enabling}, defended 1 October 2021. }

\setbiblabelwidth{1000} 
\bibliographystyle{IEEEtran}
\setstretch{0.8}
\balance
\bibliography{references}

\setstretch{1.0}
\clearpage
\appendix
\section{DRAM Trends Survey}
\label{position:appendix_a}

We survey manufacturer-recommended DRAM operating parameters as specified in
commodity DRAM chip datasheets in order to understand how the parameters have
evolved over time. We extract values from 58 independent DRAM chip datasheets
from across 19 different DRAM manufacturers with datasheet publishing dates
between 1970 and 2021. \mpk{Appendix~\ref{position:appendix_b} lists each
datasheet and the details of the DRAM chip that it corresponds to. We openly
release our full dataset on GitHub~\cite{datasheetsurveygithub}, \mpm{which
provides a spreadsheet with all of the raw data used in this paper, including
each timing and current parameter value, and additional fields (e.g., clock
frequencies, package pin counts, remaining IDD values) that are not presented
here}.}

\subsection{\mpl{DRAM Access Timing Trends}}
\label{position:subsec:timing_trends}

We survey the evolution of the following four DRAM timing parameters that are
directly related to DRAM chip performance.

\begin{itemize}
    \item \emph{tRCD}: time between issuing a row command (i.e., row activation) and a column command (e.g., read) to the row.
    \item \emph{CAS Latency (or tAA)}: time between issuing an access to a given column address and the data being ready to access.  
    \item \emph{tRAS}: time between issuing a row command (i.e., row activation) and a precharge command.
    \item \emph{tRC}: time between accessing two different rows.
\end{itemize}

\begin{figure}[h]
    \centering
    \includegraphics[width=\linewidth]{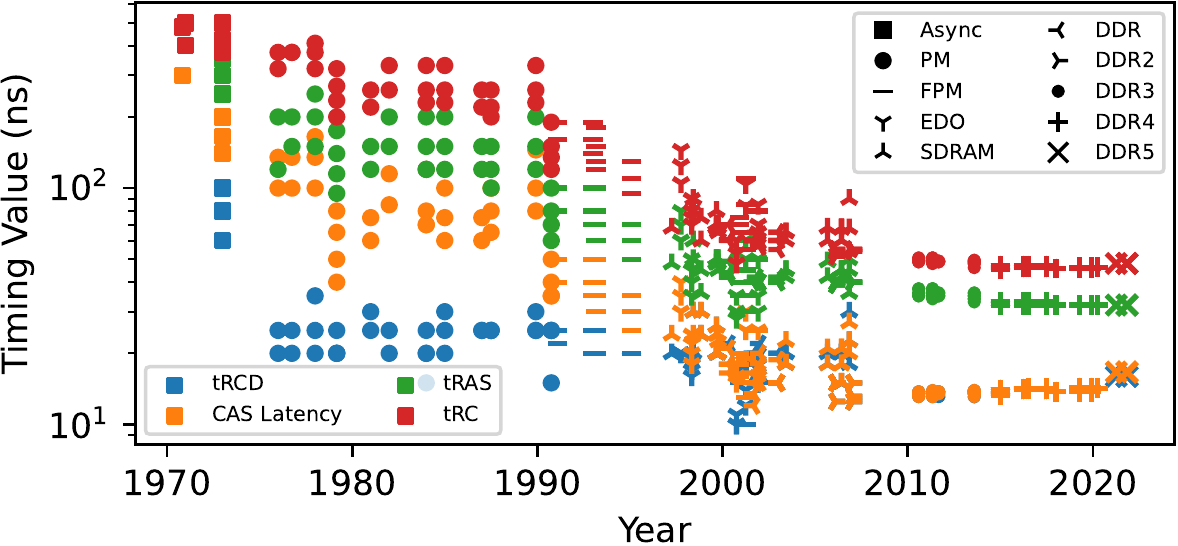}\\
    \includegraphics[width=\linewidth]{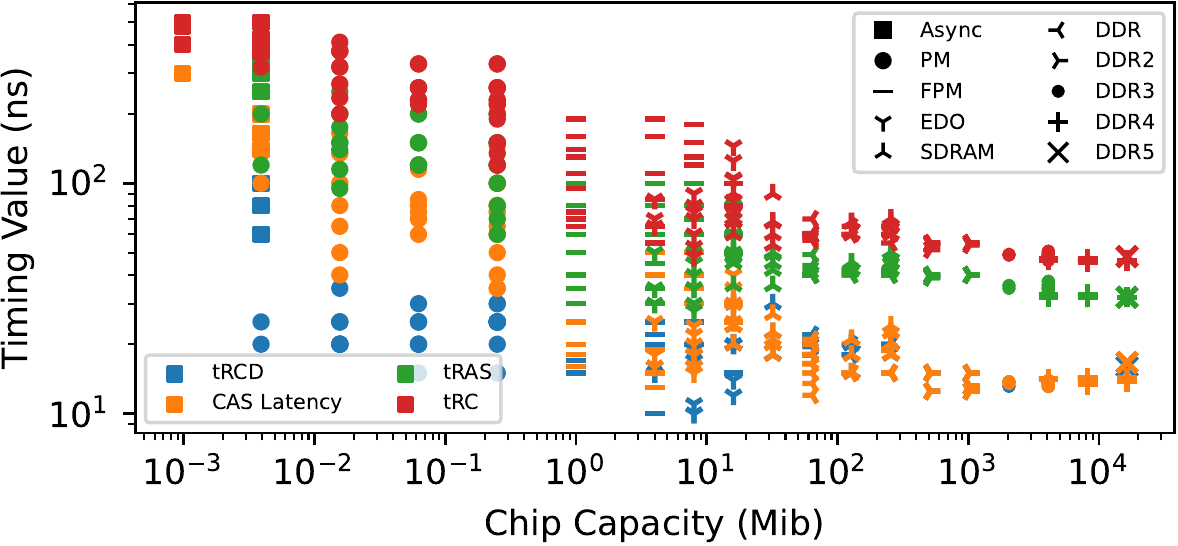}
    \caption[]{Evolution of \mpk{four} key DRAM timing parameters \mpn{(shown in
    log scale)} across years (top) and chip capacities (bottom) separated by
    DRAM standard.}
    \label{fig:da57}
\end{figure}

\noindent
Figure~\ref{fig:da57} shows how key DRAM timing parameters have evolved across
DRAM chips of different years (top) and capacities (bottom). \mpn{Timing values
are shown in log scale to better distinguish small values in newer DRAM chips.}
\mpm{Each type of marker illustrates DRAM chips of different DRAM standards.}

We make three qualitative observations. First, while all four DRAM timing values
have roughly decreased over time, improvements have been relatively stagnant for
the last two decades (note the logarithmic Y-axis). The bulk of the improvement
in timing parameter values occurred during the period of asynchronous DRAM, and
following the introduction of SDRAM and DDR\emph{n} DRAM chips, little to no
improvements have been made despite, \mpm{or possibly as a result of, continual}
increases in overall chip storage density. Second, CAS latency and tRCD
converged to roughly the same values following the introduction of synchronous
DRAM. We hypothesize that this is because \mpi{similar factors affect the
latency of these operations, including a long command and data communication
latency between the external DRAM bus and the internal storage
array~\cite{keeth2007dram}.} Third, the DDR5 data points appear to worsen
relative to previous DDR\emph{n} points. However, we believe this \mpk{might be}
because DDR5 chips are new at the time of writing this article and have not yet
been fully optimized (e.g., through die revisions and other process
improvements).

\begin{figure}[h]
    \centering
    \includegraphics[width=\linewidth]{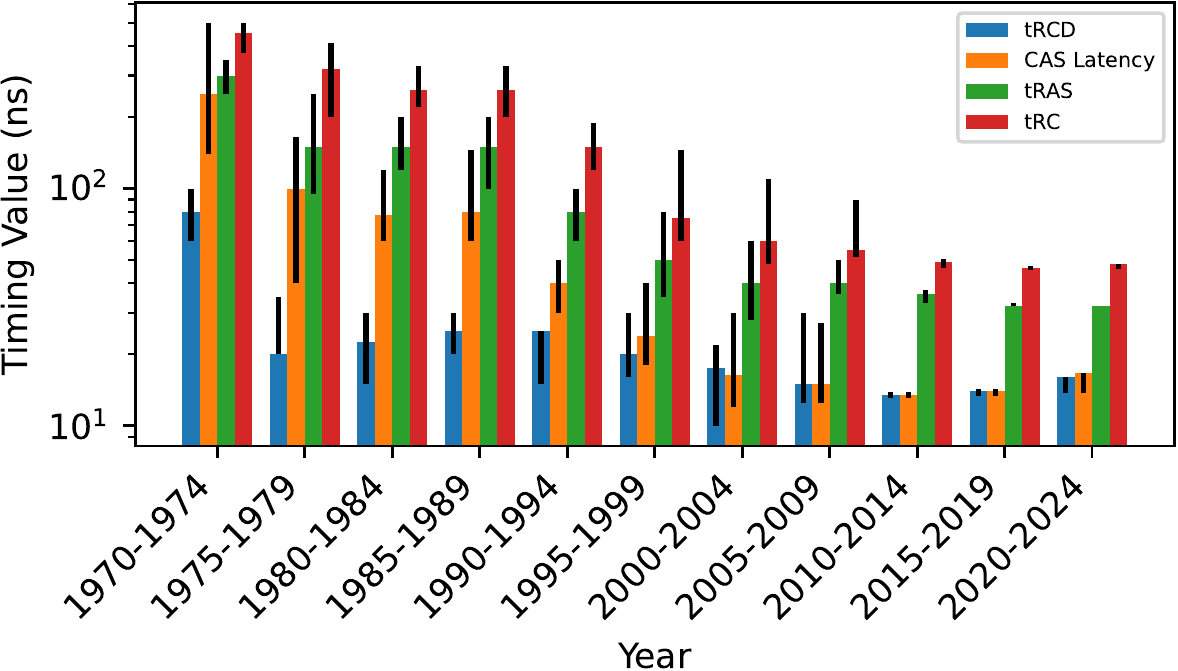}
    \includegraphics[width=\linewidth]{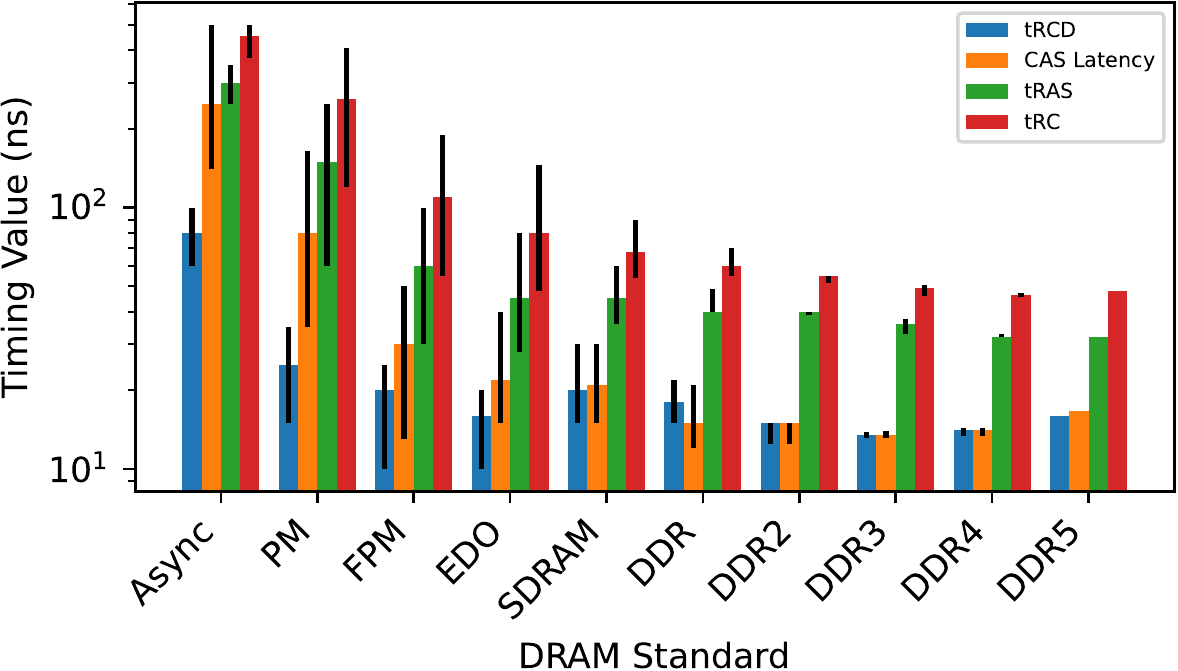}
    \caption[]{Evolution of the minimum, median, and maximum values of key DRAM timing parameters \mpn{(shown in log scale)} for each \mpk{5-year period (top) and DRAM standard (bottom)}.}
    \label{fig:da01}
\end{figure}

To quantify the changes in \mpn{access timings}, we aggregate the data points
from Figure~\ref{fig:da57} \mpm{by three different categories: time, DRAM
standard, and chip capacity. Figure~\ref{fig:da01}, shows} the minimum, median,
and maximum of the timing parameter values \mpn{(in log scale)} for
\mpk{each 5-year period (top) and DRAM standard (bottom)}. \mpk{The data shows
that the median tRCD/CAS Latency/tRAS/tRC reduced by 2.66/3.11/2.89/2.89\% per
year on average between 1970 and 2000 but only 0.81/0.97/1.33/1.53\% between
2000 and 2015\footnote{We omit the 2020 data point because 2020 shows a
regression in CAS latency due to first-generation DDR5 chips, which we believe
is not representative because of its immature technology.} for an overall
decrease of 1.83/2.10/1.99/2.00\% between 1970 and 2015.}

\begin{figure*}[b]
    \centering
    \includegraphics[width=\linewidth]{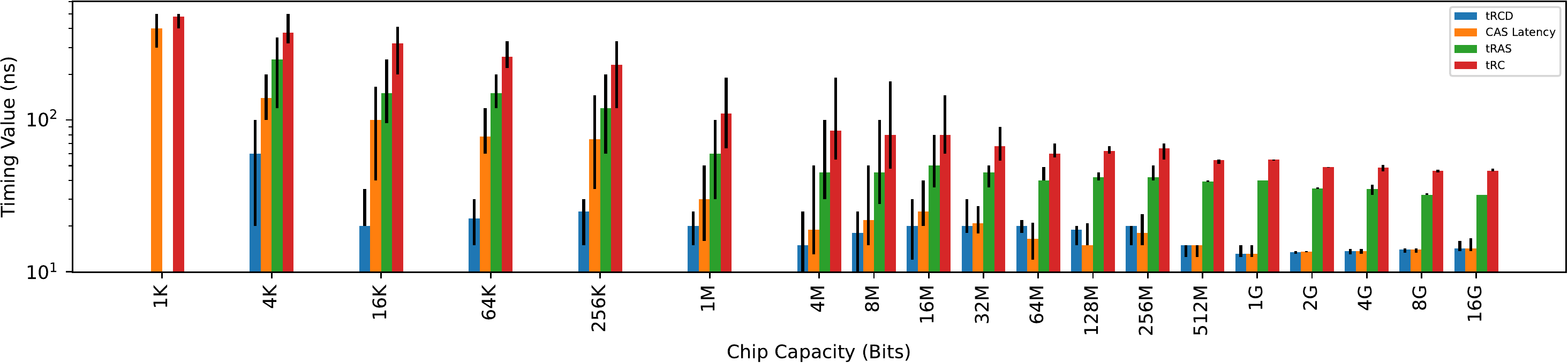}
    \caption[]{\mpm{Evolution of the minimum, median, and maximum values of key DRAM timing parameters \mpn{(shown in log scale) grouped by} DRAM chip \mpn{storage} capacity.}}
    \label{fig:da5}
\end{figure*}

\mpm{Figure~\ref{fig:da5} shows the minimum, median, and maximum of the timing
parameter values \mpn{(in log scale) grouped by DRAM chip storage
capacity.}\footnote{\mpm{We omit tRCD and tRAS for the 1~Kib chips because they
do not use a row address strobe (RAS) signal.\vspace{-1ex}}} We find that the \mpn{timings
follow similar trends as in Figure~\ref{fig:da01} because higher-capacity DRAM
chips are typically introduced more recently and follow newer} DRAM standards.}

\subsection{Current Consumption Trends}

We review the evolution of the following key DRAM current consumption
measurements, which are standardized by JEDEC and are provided by manufacturers
in their datasheets.
\begin{itemize}
    \item \emph{IDD0}: current consumption with continuous row activation and
    precharge commands issued to only one bank.
    \item \emph{IDD4R}: current consumption when issuing back-to-back read operations to all banks.
    \item \emph{IDD5B}: current consumption when issuing continuous burst refresh operations.
\end{itemize}

\begin{figure}[H]
    \centering
    \includegraphics[width=\linewidth]{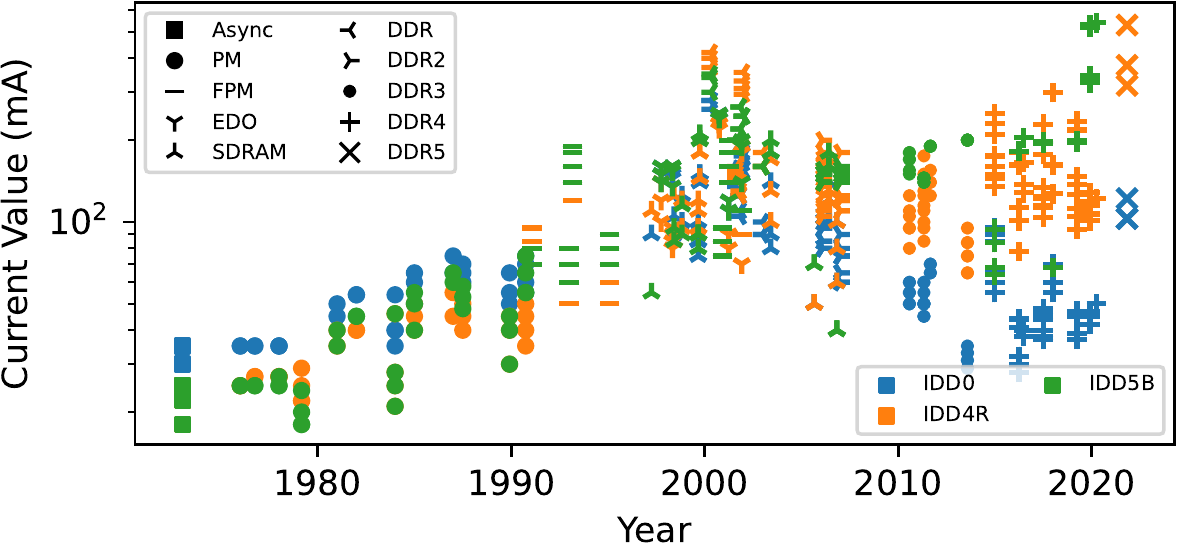}
    \includegraphics[width=\linewidth]{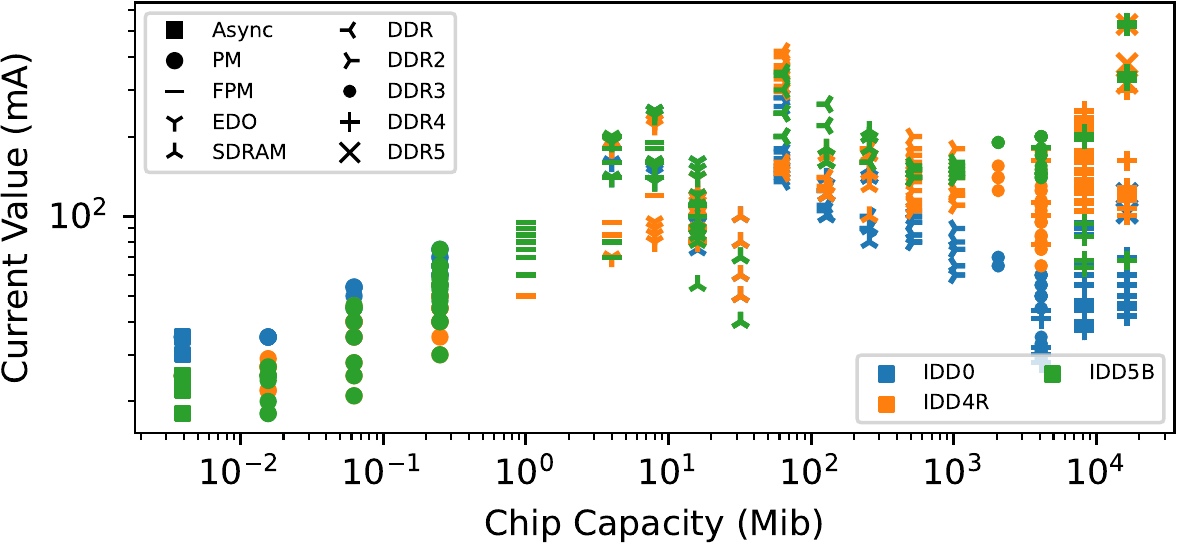}
    \caption[]{Evolution of key DRAM current consumption values \mpn{(shown in log scale)} across years (top) and chip capacities (bottom) separated by DRAM standard.}
    \label{fig:da46}
\end{figure}

\noindent
Figure~\ref{fig:da46} shows how key DRAM current consumption values \mpn{(in log
scale)} have evolved across DRAM chips of different years (top) and capacities
(bottom). We use different markers to show data points from chips of different
DRAM standards. \mpn{We qualitatively observe that current consumption increased
exponentially up until approximately the year 2000, which is about the time at
which improvements in access timings slowed down (as seen in
Figure~\ref{fig:da57}). After this point, different current consumption
measurements diverged as IDD0 values decreased while IDD4R and IDD5B stabilized
or increased. We explain this behavior by a change in the way DRAM chips were
refreshed as DRAM capacities continued to increase. Earlier DRAM chips refreshed
rows using individual row accesses (e.g., RAS-only refresh), which result in
comparable behavior for access and refresh operations. In contrast, newer DRAM
chips aggressively refresh \emph{multiple} rows per refresh operation (e.g.,
burst refresh), which differentiates refresh operations from normal row
accesses~\cite{mukundan2013understanding,balasubramonian2019innovations,
utah2013dram}.} 

\begin{figure}[h]
    \centering
    \includegraphics[width=\linewidth]{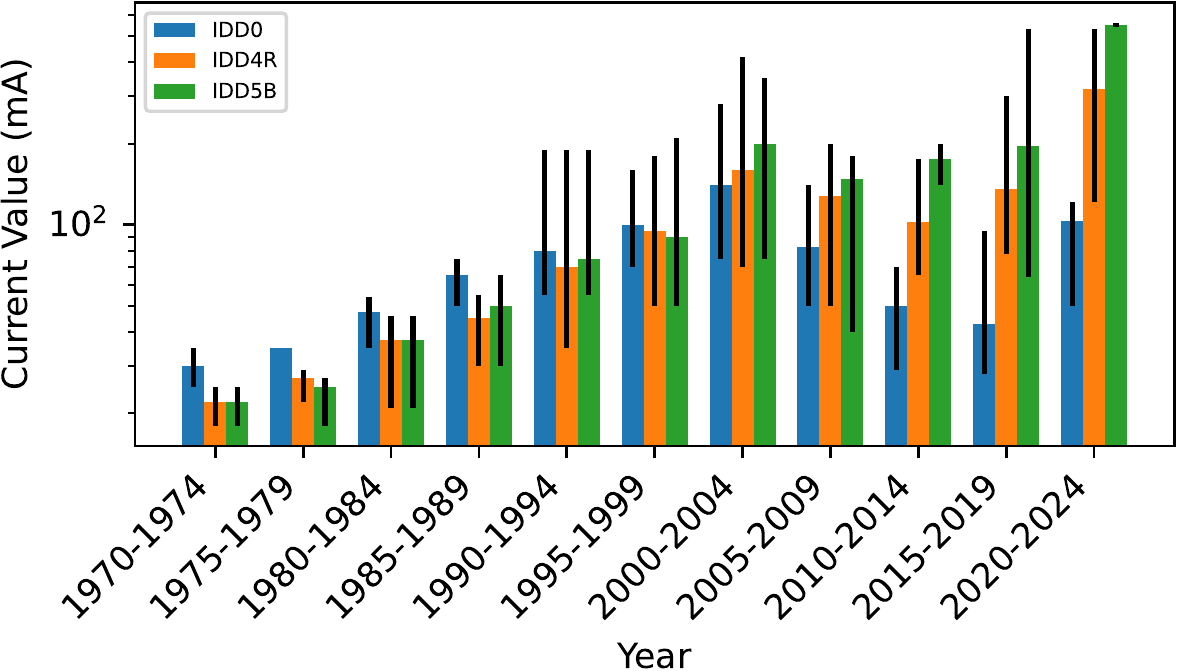}
    \includegraphics[width=\linewidth]{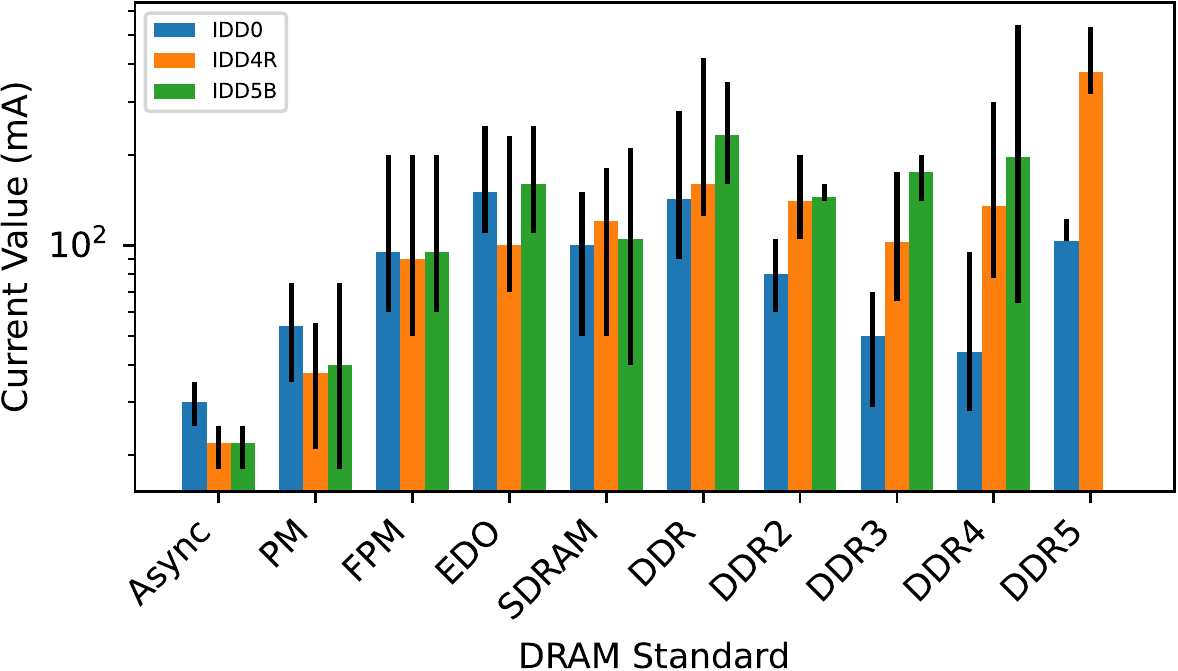}
    \caption[]{Evolution of the minimum, median, and maximum of key DRAM current consumption value \mpn{(shown in log scale)} for each \mpk{5-year period (top) and DRAM standard (bottom).}}
    \label{fig:da23}
\end{figure}

We quantify the current \mpn{consumption} values by aggregating the data points
from Figure~\ref{fig:da46} \mpm{by time and DRAM standard.}
Figure~\ref{fig:da23} shows \mpm{the minimum, median, and maximum values
\mpn{(in log scale)} across each} \mpk{5-year period (top) and DRAM standard
(bottom).} \mpk{The data shows that the median IDD0/IDD4R/IDD5B increased by
12.22/20.91/26.97\% per year on average between 1970 and 2000 but
\emph{decreased} by 4.62/1.00/0.13\% between 2000 and 2015\footnote{Similar to
Section~\ref{position:subsec:timing_trends}, we omit the 2020 data point because
the first-generation DDR5 chips exhibit outlying data values (e.g., no data
reported for IDD5B in the datasheets).} for an overall increase of
0.96/11.5/17.5\% between 1970 and 2015.}

\subsection{Relationship Between Timings and Currents}

\begin{figure}[b]
    \centering
    \includegraphics[width=\linewidth]{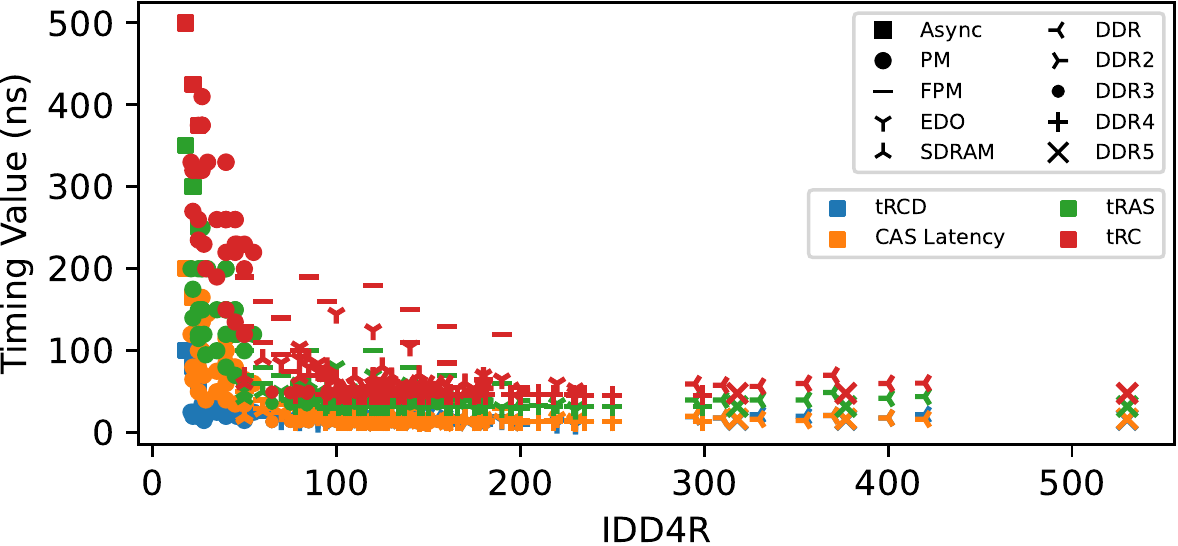}
    \caption[]{\mpk{Relationship between the four timing parameters and IDD4R separated by DRAM standard.}}
    \label{fig:da9}
\end{figure}

Finally, we examine the high-level relationship between the timing parameter and
current consumption values. We find that the two are generally inversely
related, which follows from the general principle that faster DRAM chips (i.e.,
lower timing parameters) require more power (i.e., increased current consumption
values). Figure~\ref{fig:da9} illustrates this relationship for the four timing
parameters studied in Section~\ref{position:subsec:timing_trends} relative to
IDD4R (i.e., the current consumption of read operations).

\subsection{\mpl{DRAM Refresh Timing Trends}}

\mpl{DRAM refresh is governed by two key timing parameters:}

\begin{itemize}
    \item \mpl{\emph{tREFI} (refresh interval): time between consecutive refresh commands sent by the memory controller.}
    \item \mpl{\emph{tRFC}: duration of a single refresh command.}
\end{itemize}

\begin{figure}[h]
    \centering
    \includegraphics[width=\linewidth]{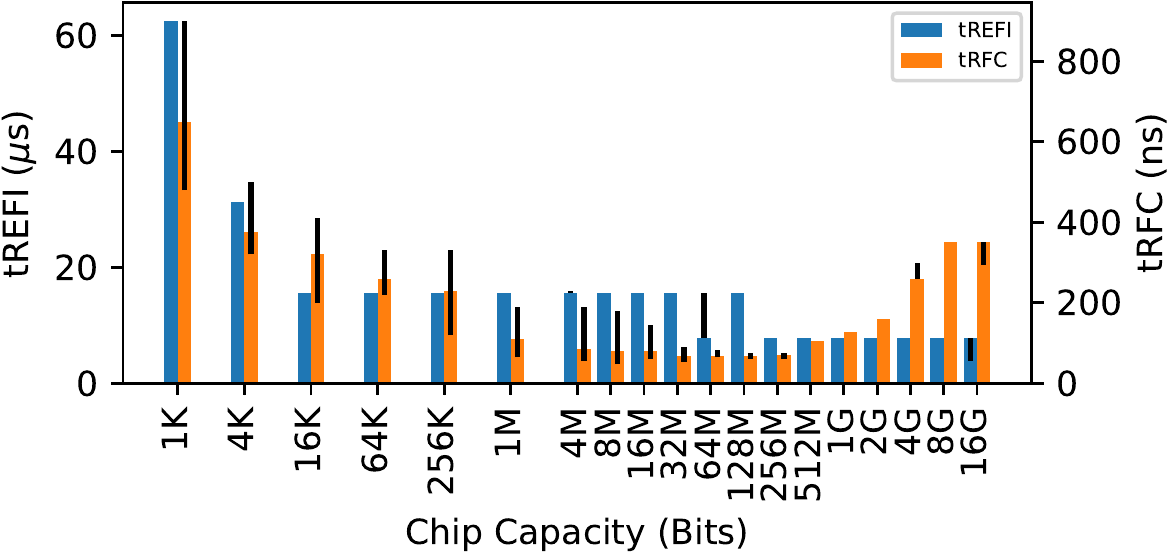}
    \caption[]{Evolution of tREFI (left y-axis) and tRFC (right y-axis) across DRAM chips of increasing storage capacity.}
    \label{fig:da12}
\end{figure}

\noindent
\mpl{Figure~\ref{fig:da12} shows how tREFI (left y-axis) and tRFC (right y-axis)
evolved across the DRAM chips in our study. We group chips by storage capacity
because DRAM refresh timings are closely related to capacity:
\mpm{higher-capacity} chips using the same technology require more time or more
refresh operations to fully refresh. \mpm{The error bars show the minimum and
maximum values observed across all chips for any given chip capacity.}}

\mpl{We make three observations. First, tREFI is shorter for
\mpm{higher-capacity} DRAM chips (e.g., 62.5~$\mu$s for an asynchronous 1~Kib
chip versus 3.9~$\mu$s for a 16~Gib DDR5 chip). This is consistent with the fact
that \mpm{higher-capacity} chips require more frequent refreshing. Second, tRFC
first decreases with chip capacity (e.g., 900 ns for an asynchronous 1~Kib chip
versus 54~ns for a 32~Mib SDRAM chip) but then increases (e.g., to 350~ns for a
16~Gib DDR4 chip). This is because rapid improvements in row access times (and
therefore refresh timings) initially outpaced the increase in storage capacity.
However, starting around 512~Mib chip sizes, row access times improved much more
slowly (as observed in Section~\ref{position:subsec:timing_trends}) while
storage capacity continued to increase. \mpm{Third, the variation in tRFC across
chips of each capacity (illustrated using the error bars) decreased for
\mpm{higher-capacity} chips. This is because higher-capacity chips follow more
recent DRAM standards (i.e., DDR\emph{n}), which standardize DRAM auto refresh
timings. In contrast, older DRAM chips were simply refreshed as quickly as their
rows could be accessed (e.g., every tRC using RAS-only refresh).}}

\begin{figure}[h]
    \centering
    \includegraphics[width=\linewidth]{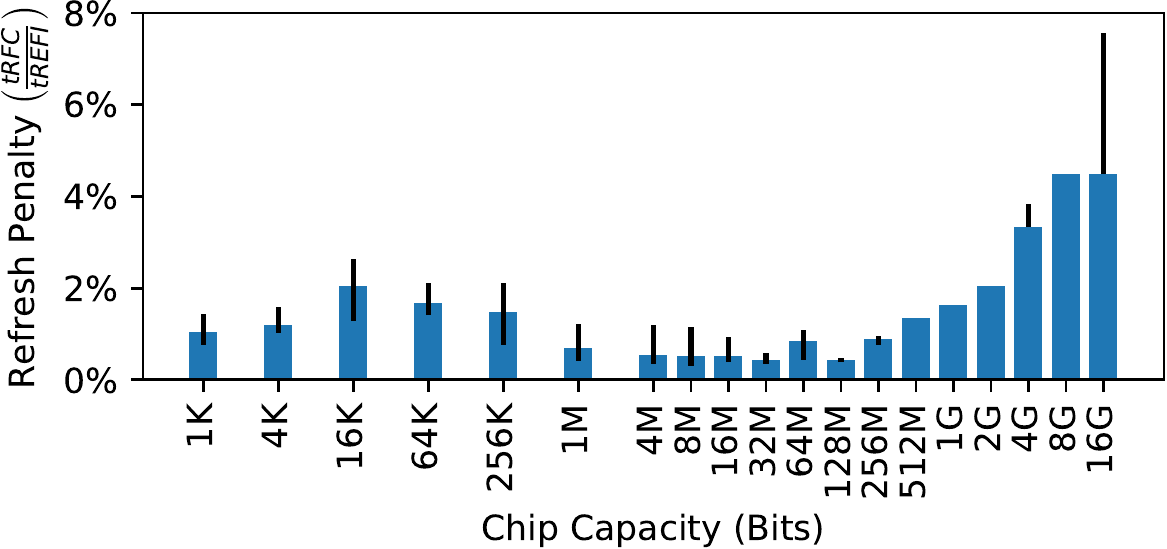}
    \caption[]{\mpm{Refresh penalty} (computed as the ratio between tRFC and tREFI) for DRAM chips of different storage capacities.}
    \label{fig:da_ref_overhead}
\end{figure}

\mpm{Figure~\ref{fig:da_ref_overhead} shows the \emph{refresh
penalty}~\cite{utah2013dram, balasubramonian2019innovations}, which is defined
as the ratio between tRFC and tREFI, for DRAM chips of different storage
capacities. The refresh penalty represents the average time that a DRAM rank (or
bank) is unavailable for access due to refresh operations~\cite{utah2013dram,
stuecheli2010elastic, balasubramonian2019innovations, cheng2019retention,
zhang2014cream}. We observe that the refresh penalty exhibits a similar trend to
tRFC: the refresh penalty worsens from a median of 1.04\% for 1~Kib chips to
2.05\% for 16~Kib chips, then improves to 0.43\% for 128~Mib chips, and finally
worsens to a median of 4.48\% (worst-case of 7.56\% for DDR5 chips) for 16~Gib
chips.}

\mpl{This non-monotonic trend is due to the relative improvements in DRAM access
times and storage capacities: DRAM capacities consistently improved while DRAM
access times did so only for older, lower-capacity chips (e.g., $\le 128$~Mib
chips). This is consistent with trends observed in prior
work~\mpm{\cite{nguyen2018nonblocking, liu2012raidr, bhati2015flexible,
nair2013case, chang2014improving, qureshi2015avatar}}, which expect that future,
higher-capacity DRAM chips will spend an even larger proportion of time
refreshing unless the DRAM refresh algorithm \mpm{and techniques} can be
improved.}

\clearpage
\onecolumn

\section{\mpk{Survey Data Sources}}
\label{position:appendix_b}
\mpk{Table~\ref{position:tab:sources} itemizes the 58 \mpm{DRAM} datasheets used
for our survey in Appendix~\ref{position:appendix_a}. For each datasheet, we
show the DRAM chip manufacturer, model number, DRAM standard, year, and
capacity. Our full dataset is available online~\cite{datasheetsurveygithub}.}

\begin{center}
    \small
    \renewcommand{\arraystretch}{0.90}
    \begin{tabular}{llllll}
        \textbf{Year} & \textbf{Manufacturer} & \textbf{Model Number} & \textbf{Datasheet Source} & \textbf{DRAM Standard} & \textbf{Capacity per Chip (Kib)} \\\hline
        1970 & Intel             & 1103         & \citeS{ds_intel19701103}            & Asynchronous & 1 \\
        1971 & Mostek            & MK4006       & \citeS{ds_mostek1971mk4006}         & Asynchronous & 1 \\
        1973 & Mostek            & MK4096       & \citeS{ds_mostek1973mk4096}         & Asynchronous & 4 \\
        1976 & Mostek            & MK4027       & \citeS{ds_mostek1976mk4027}         & PM           & 4 \\
        1976 & Mostek            & MK4116P      & \citeS{ds_mostek1976mk4116p}        & PM           & 16 \\
        1978 & Fairchild         & F4116        & \citeS{ds_fairchild1978f4116}       & PM           & 16 \\
        1979 & Intel             & 2118         & \citeS{ds_intel19792118}            & PM           & 16 \\
        1981 & Mitsubishi        & M5K4164ANP   & \citeS{ds_mitsubishi1981m5k4164anp} & PM           & 64 \\
        1982 & Mostek            & MK4564       & \citeS{ds_mostek1982mk4564}         & PM           & 64 \\
        1984 & NTE               & NTE4164      & \citeS{ds_nte1984nte4164}           & PM           & 64 \\
        1984 & Texas Instruments & TMS4416      & \citeS{ds_texas1984tms4416}         & PM           & 64 \\
        1985 & Mitsubishi        & M5M4256P     & \citeS{ds_mitsubishi1985m5m4256p}   & PM           & 256 \\
        1987 & Samsung           & KM41464A     & \citeS{ds_samsung1987km41464a}      & PM           & 256 \\
        1987 & Texas Instruments & TMS4464      & \citeS{ds_texas1987tms4464}         & PM           & 256 \\
        1989 & Texas Instruments & SMJ4464      & \citeS{ds_texas1989smj4464}         & PM           & 256 \\
        1990 & Intel             & 21256        & \citeS{ds_intel199021256}           & PM           & 256 \\
        1991 & Mitsubishi        & M5M44100     & \citeS{ds_mitsubishi1991m5m44100}   & FPM          & 4096 \\
        1993 & Mitsubishi        & M5M44256B    & \citeS{ds_mitsubishi1993m5m44256b}  & FPM          & 1024 \\
        1993 & Mosel Vitelic     & V404J8       & \citeS{ds_mosel1993v404j8}          & FPM          & 8192 \\
        1995 & Siemens           & HYB511000BJ  & \citeS{ds_siemens1995hyb511000bj}   & FPM          & 1024 \\
        1997 & Hyundai           & HY5118164B   & \citeS{ds_hyundai1997hy5118164b}    & EDO          & 16384 \\
        1997 & Samsung           & KM48S2020CT  & \citeS{ds_samsung1997km48s2020ct}   & SDRAM        & 16384 \\
        1998 & Micron            & MT48LC4M4A1  & \citeS{ds_micron1998mt48lc4m4a1}    & SDRAM        & 16384 \\
        1998 & Mosel Vitelic     & V53C808H     & \citeS{ds_mosel1998v53c808h}        & EDO          & 8192 \\
        1998 & Siemens           & HYB39S16400  & \citeS{ds_siemens1998hyb39s16400}   & SDRAM        & 16384 \\
        1999 & Samsung           & K4S160822D   & \citeS{ds_samsung1999k4s160822d}    & SDRAM        & 16384 \\
        1999 & Samsung           & K4S561632A   & \citeS{ds_samsung1999k4s561632a}    & SDRAM        & 262144 \\
        2000 & Amic              & A416316B     & \citeS{ds_amic2000a416316b}         & FPM          & 1024 \\
        2000 & ISSI              & IS41LV32256  & \citeS{ds_issi2000is41lv32256}      & EDO          & 8192 \\
        2000 & Samsung           & K4D623237A5  & \citeS{ds_samsung2000k4d623237a5}   & DDR          & 65536 \\
        2001 & Alliance          & AS4C256K16E0 & \citeS{ds_alliance2001as4c256k16e0} & EDO          & 4096 \\
        2001 & Alliance          & AS4C4M4FOQ   & \citeS{ds_alliance2001as4c4m4foq}   & FPM          & 16384 \\
        2001 & ISSI              & IS41C4400X   & \citeS{ds_issi2001is41c4400x}       & EDO          & 16384 \\
        2001 & Micron            & MT46V2M32    & \citeS{ds_micron2001mt46v2m32}      & DDR          & 65536 \\
        2001 & Micron            & MT46V32M4    & \citeS{ds_micron2001mt46v32m4}      & DDR          & 131072 \\
        2001 & Mosel Vitelic     & V58C265164S  & \citeS{ds_mosel2001v58c265164s}     & DDR          & 65536 \\
        2001 & TM Tech           & T224160B     & \citeS{ds_tm2001t224160b}           & FPM          & 4096 \\
        2003 & Micron            & MT46V64M4    & \citeS{ds_micron2003mt46v64m4}      & DDR          & 262144 \\
        2003 & Samsung           & K4S560432E   & \citeS{ds_samsung2003k4s560432e}    & SDRAM        & 262144 \\
        2005 & Amic              & A43L0632     & \citeS{ds_amic2005a43l0632}         & SDRAM        & 32768 \\
        2006 & Elite             & M52S32321A   & \citeS{ds_elite2006m52s32321a}      & SDRAM        & 32768 \\
        2006 & ISSI              & IS42S81600B  & \citeS{ds_issi2006is42s81600b}      & SDRAM        & 131072 \\
        2006 & Sasmung           & K4T51043QC   & \citeS{ds_sasmung2006k4t51043qc}    & DDR2         & 524288 \\
        2007 & Micron            & MT47H256M4   & \citeS{ds_micron2007mt47h256m4}     & DDR2         & 1048576 \\
        2010 & Samsung           & K4B4G0446A   & \citeS{ds_samsung2010k4b4g0446a}    & DDR3         & 4194304 \\
        2011 & Hynix             & H5TQ4G43MFR  & \citeS{ds_hynix2011h5tq4g43mfr}     & DDR3         & 4194304 \\
        2011 & Nanya             & NT5CB512M    & \citeS{ds_nanya2011nt5cb512m}       & DDR3         & 2097152 \\
        2013 & Samsung           & K4B4G0446A   & \citeS{ds_samsung2013k4b4g0446a}    & DDR3         & 4194304 \\
        2015 & Micron            & MT40A2G      & \citeS{ds_micron2015mt40a2g}        & DDR4         & 8388608 \\
        2016 & Hynix             & H5AN4G4NAFR  & \citeS{ds_hynix2016h5an4g4nafr}     & DDR4         & 4194304 \\
        2016 & Samsung           & K4A8G165WC   & \citeS{ds_samsung2016k4a8g165wc}    & DDR4         & 8388608 \\
        2017 & Hynix             & H5AN8G4NAFR  & \citeS{ds_hynix2017h5an8g4nafr}     & DDR4         & 8388608 \\
        2018 & Micron            & MT40A        & \citeS{ds_micron2018mt40a}          & DDR4         & 16777216 \\
        2019 & Hynix             & H5AN8G4NCJR  & \citeS{ds_hynix2019h5an8g4ncjr}     & DDR4         & 8388608 \\
        2019 & Samsung           & K4AAG045WA   & \citeS{ds_samsung2019k4aag045wa}    & DDR4         & 16777216 \\
        2020 & Samsung           & K4AAG085WA   & \citeS{ds_samsung2020k4aag085wa}    & DDR4         & 16777216 \\
        2021 & Hynix             & HMCG66MEB    & \citeS{ds_hynix2021hmcg66meb}       & DDR5         & 16777216 \\
        2021 & Micron            & MT60B1G16    & \citeS{ds_micron2021mt60b1g16}      & DDR5         & 16777216
    \end{tabular}
\end{center}
\captionof{table}{List of DRAM chip datasheets used in our DRAM trends survey.}
\label{position:tab:sources}

\clearpage
\twocolumn
\balance
\bibliographystyleS{IEEEtran}
\setstretch{0.8}
\bibliographyS{references}

\end{document}